\documentclass[11pt]{article}
\pdfoutput=1
\usepackage{amsfonts,amsmath,amssymb,authblk,color,enumerate,float,graphicx,multicol,multirow,placeins,pifont,slashed,titlesec,xspace}
\usepackage[labelfont=bf]{caption}
\usepackage[top=1in,bottom=1in,left=1in,right=1in,footskip=0.5in,headheight=0.5in,footnotesep=0.2in,marginparwidth=0in,marginparsep=0in]{geometry}
\usepackage{jcappub}
\usepackage{aas_macros}
\usepackage{hyphenat}
\usepackage{enumitem}

\long\def\symbolfootnote[#1]#2{\begingroup%
\def\thefootnote{\fnsymbol{footnote}}\footnote[#1]{#2}\endgroup}

\newcommand\notype[1]{\unskip}

\newcommand{\anr}{a_{ \textrm{\tiny{NR}}}}

\newcommand{\rhoDM}{\rho_{\textrm{\tiny{DM}}}}

\newcommand{\TSM}{T_{\textrm{SM}}}
\newcommand{\aeq}{a_{\textrm{eq}}}
\newcommand{\sigmaDM}{\sigma_{\textrm{DM}}}
\newcommand{\deltaDM}{\delta_{\textrm{DM}}}
\newcommand{\thetaDM}{\theta_{\textrm{DM}}}

\newcommand{\Mpc}{\textrm{Mpc}}
\newcommand{\cm}{\textrm{cm}}
\newcommand{\g}{\textrm{g}}

\newcommand{\usq}{\left<u^2\right>}

\newcommand{\keV}{\textrm{keV}}

\newcommand{\Lya}{Lyman-$\alpha$ }

\title{The Cosmological Evolution of Self-interacting Dark Matter}
\author{Daniel Egana-Ugrinovic$^{1}$,}
\author{Rouven Essig$^{2}$,}
\author{Daniel Gift$^{2}$,}
\author{and Marilena LoVerde$^{2}$}
\affiliation{
$^{1}$Perimeter Institute for Theoretical Physics, Waterloo, ON, N2L 2Y5\\
$^{2}$C. N. Yang Institute for Theoretical Physics, Stony Brook University, 
 Stony Brook, NY 11794\\
}

 \abstract{
We study the evolution of cosmological perturbations in dark-matter models with elastic and velocity-independent self interactions.
Such interactions are imprinted in the matter-power spectrum as dark acoustic oscillations,
which can be experimentally explored to determine the strength of the self scatterings. 
Models with self interactions have similarities to warm dark matter, as they lead to suppression of power on small scales when the dark-matter velocity dispersion is sizable. 
Nonetheless, both the physical origin and the extent of the suppression differ for self-interacting dark matter from conventional warm dark matter, 
with a dark sound horizon controlling the reduction of power in the former case, and a free-streaming length in the latter.
We thoroughly analyze these differences by performing computations of the linear power spectrum using a newly developed Boltzmann code.
We find that 
while current \Lya data disfavor conventional warm dark matter with a mass less than 5.3\,\keV, when self interactions are included at their maximal value consistent with bounds from the Bullet Cluster, the limits are relaxed to 4.4\,\keV. 
Finally, we make use of our analysis to set novel bounds on light scalar singlet dark matter.
\\[0.2cm]

\noindent
\textit{Preprint: YITP-SB-2020-39}
}


\begin{document}
\maketitle
\flushbottom
\newpage

\newpage
\section{Introduction}

The search for dark matter in laboratories and astronomical observatories via its interactions with the Standard Model (SM) is ongoing, but all evidence for its existence remains purely gravitational in nature. While theoretical motivations exist for the dark matter to have non-gravitational interactions with the SM, it is also entirely possible that the dark sector remains secluded and interacts with the visible sector purely gravitationally, or via experimentally inaccessible non-gravitational interactions.
A secluded dark sector can be successfully populated in the early universe via a variety of mechanisms that do not require interactions with the SM, including gravitational production~\cite{Parker:1969au,PhysRevD.22.322,Ford:1986sy}, interactions with the inflaton~\cite{Kofman:1994rk,Adshead:2016xxj}, or the misalignment mechanism~\cite{Dine:1982ah,Abbott:1982af,preskill1983cosmology}.
It is possible, however, that the dark sector has non-gravitational dynamics of its own, both due to theoretical motivations and experimental hints, which suggest that dark interactions are responsible for the anomalous behavior of cosmic structure on small scales~\cite{Spergel:1999mh}. 

If this is the case, progress in our understanding of the dark sector can still be made by looking for the signatures that these interactions leave on gravitationally bound visible matter.
Depending on the complexity of the dark interactions, a variety of phenomenological signatures arise, including modifications of halo shapes~\cite{Peter:2012jh}, large-scale acoustic oscillations~\cite{deLaix:1995vi,CyrRacine:2012fz}, and even the formation of complete dark galaxies~\cite{Chang:2018bgx,Fan:2013yva}. 
Even the simplest of the dynamical dark sector theories, namely a dark matter particle with elastic, isotropic, and velocity-independent self interactions~\cite{Carlson:1992fn}, leaves visible signs on gravitationally bound matter. 
In what follows, we refer to this model as self-interacting dark matter (SIDM). 
Observable signatures left by SIDM on cosmic structure are, \textit{e.g.}, reducing the offsets between the dark and visible gas distributions when clusters of galaxies encounter~\cite{Markevitch:2003at} and giving a spherical shape to otherwise triaxial halos~\cite{Peter:2012jh}.
SIDM has also been proposed to solve a variety of small-scale problems, such as the explanation of cores in dwarf galaxies~\cite{Spergel:1999mh,deBlok:2009sp,Rocha:2012jg,zavala2013constraining,Elbert:2014bma,Kaplinghat:2015aga,Bullock:2017xww,Tulin:2017ara,Kaplinghat:2019bho}.

In this work we explore an additional signature left by SIDM on cosmic structure, 
which is the imprint of acoustic oscillations on the matter-power spectrum. 
The effect of SIDM on the matter-power spectrum is similar to the one of conventional non-interacting warm-dark matter (WDM),  
as in both models power is reduced on scales below a characteristic cutoff, controlled mostly by the particle's velocity dispersion~\cite{deLaix:1995vi}.
In SIDM, the cutoff scale is the dark sound horizon due to pressure support, while in WDM, it corresponds to the free-streaming length. 
To understand the subtle differences between the two cases, 
we investigate in detail the whole range of self-scattering cross sections starting from $\sigma/m=0$, as in models of WDM, and going up to the maximal values allowed by the Bullet Cluster and other probes, $\sigma/m \sim 1\, \cm^2/\g$~\cite{Randall:2007ph}. 
In order to do so, we develop a new Boltzmann code to compute the linear power spectrum,
taking into account the cosmic evolution  of acoustic oscillations, decoupling of the self interactions, and the later periods of free-streaming.
We thoroughly check and validate our code by comparing its results with the Boltzmann solver CLASS \cite{Blas_2011} for the special cases of cold and warm dark matter with no self interactions (where CLASS can be used), finding sub-percent level precision in our computations.

The imprints left by SIDM or WDM on the linear matter-power spectrum can be observationally investigated and distinguished from each other via precise measurements of cosmic structure on small scales,
such as the ones provided by current and/or future measurements of the \Lya forest~\cite{Irsic:2017ixq,PhysRevD.98.083540}, Milky-Way satellite counts~\cite{PhysRevD.83.043506,jethwa2018upper,kennedy2014constraining},
strong lensing~\cite{Birrer_2017,gilman2019,gilman2019_newest} stellar streams~\cite{banik2019novel,Banik_2018_naturedm,Dalal:2020mjw}, and high-resolution lensing measurements of the cosmic microwave background (CMB)~\cite{Nguyen:2017zqu,Sehgal:2019ewc}.  
In particular, using such probes and calculations of the matter-power spectrum for a given dark sector model, 
bounds can be set on the corresponding dark matter mass or velocity dispersion. 
Here we will focus on calculating \Lya bounds on SIDM, 
but a large part of our analysis (especially our linear power-spectrum computations) can also be used for studying other probes.  
From our analysis, we find that bounds from \Lya on the dark matter mass slightly decrease as the self-interaction cross section is increased.
More specifically, 
we show that while current \Lya bounds disfavor typical warm dark matter\footnote{As a convention, the benchmark WDM model corresponds to a single non-interacting Weyl fermionic species with a vanishing primordial chemical potential~\cite{1986ApJ...304...15B}.} for masses $m\leq 5.3\,\keV$~\cite{Irsic:2017ixq} ($m\leq 3.5\,\keV$ under more conservative assumptions for current \Lya bounds), 
when large self interactions are allowed, $\sigma/m=1\,\cm^2/\g$, 
the bound decreases to $m\leq 4.4\,\keV$ ($m\leq 2.95 \,\keV$ for conservative bounds). 

In terms of concrete particle physics models, our results set stringent and previously unexplored constraints on one of the simplest dark-matter models: scalar singlet dark matter. 
In particular, we show that for singlet-self couplings in the range $5\times 10^{-8}\lesssim \lambda \lesssim 10^{-5}$ (where the upper range is imposed for consistency with Bullet Cluster bounds~\cite{Randall:2007ph}) and for a present dark-matter velocity dispersion $\sqrt{\usq/3} \gtrsim 10^{-8}$, 
observable acoustic oscillations are imprinted in the linear matter-power spectrum in this model.
For couplings $ \lambda \lesssim 5\times 10^{-8}$, free-streaming affects modes relevant for \Lya instead. 
Both the effects of pressure support and free-streaming are constrained by our analysis. 
Concretely, our results improve previous constraints on scalar singlet dark matter (which come from the effective number of relativistic species $N_{eff}$~\cite{Arcadi:2019oxh}) by orders of magnitude in mass or velocity dispersion.

Given the large amount of literature written on SIDM, it is convenient to briefly comment about previous analyses related to our work. 
To the best of our knowledge, the earliest reference studying the effect of self interactions in the linear matter-power spectrum is~\cite{deLaix:1995vi}, and further analysis was carried out in~\cite{AtrioBarandela:1996ur,Hannestad:2000gt}.
In these references some of the main effects that we study in this work were laid out.
Here we improve on these analyses by performing a comprehensive and detailed study of the cosmological evolution of SIDM for all the allowed range of self-interaction cross sections, and also obtain updated bounds from Lyman-$\alpha$.
More recently, self interactions have been studied in the context of ETHOS~\cite{Cyr-Racine:2015ihg,Bose:2018juc,Archidiacono:2019wdp}, 
which however concentrates mostly on specific benchmarks for velocity-dependent cross sections.
\Lya bounds on self interactions are also analyzed in~\cite{Huo:2019bjf}, but that reference does not study late-time suppression on small scales due to acoustic oscillations.
Reference~\cite{Yunis:2020woq} presents a detailed study of the Boltzmann hierarchies in SIDM at the theoretical level, while~\cite{March-Russell:2020nun,Heimersheim:2020aoc,Das:2018ons} studied models with self-interactions and briefly comment on bounds from Lyman-$\alpha$ and others.
In~\cite{Garny:2018byk}, the authors present approximate constraints on strongly self-interacting dark matter from \Lya assuming the constraints are equal to the ones on WDM.
Finally, we do not study quantum effects that arise in self-interacting theories when the dark-matter particle is ultra-light, but 
we refer the reader to~\cite{Hu:2000ke,Arvanitaki:2014faa} for the corresponding discussion. 

We organize this paper as follows.
In section \ref{sec:background}, we discuss the background evolution of the homogeneous component of self-interacting dark matter. 
In section \ref{sec:cosmopert}, we discuss the dark-matter perturbations, 
the free-streaming and sound-horizon scales, 
and the evolution of the power spectrum. 
We present our results in section \ref{sec:results}, including bounds from Lyman-$\alpha$. 
We then move to discuss the singlet-scalar model and bounds its parameter space from \Lya probes in section \ref{sec:singlet}. 
We conclude in section \ref{sec:conclusions}.  We provide several technical details in appendices. 

\section{Background thermodynamics of self-interacting dark matter}
\label{sec:background}

The cosmological evolution of self-interacting dark matter can be divided into the evolution of the homogeneous background component
and the evolution of the small dark-matter perturbations on top of this background.
In this section, we study the basic thermodynamics and evolution of the homogeneous dark-matter background, 
which is a non-relativistic component at present times.
Throughout sections \ref{sec:equilibriumdistro}-\ref{sec:phasespace}, we discuss the background evolution and phase-space distributions of dark matter while it is in kinetic equilibrium. 
As the universe expands, 
interactions become inefficient, the dark matter falls out of equilibrium and the phase-space distributions are frozen out. 
We provide an approximate prescription for the decoupled phase-space distributions in section \ref{sec:decoupling}.
In this work, we limit ourselves to discussing models containing a single elastically self-interacting dark-matter component. 
We also consider only models with velocity-independent self-interaction cross sections. 
We do not include any interactions with baryons. 
Finally, it is worth pointing out that in the literature kinetic equilibrium/decoupling usually refer to the elastic interactions between the dark sector and some external thermal bath. 
For instance, in WIMP models, kinetic decoupling indicates when the elastic interactions between the dark sector and the SM decouple. 
In this work, however, we study a dark sector that is secluded from the SM or other thermal baths and only interacts with itself.
Thus, in what follows kinetic decoupling is used to describe when the elastic \textit{self-interactions} within the dark sector decouple.

\subsection{Phase space distributions and equilibrium thermodynamics}
\label{sec:equilibriumdistro}
The dark matter is described by its phase-space distribution $f(x,q)$, 
which determines the number of particles within a phase-space differential.
We closely follow the conventions of~\cite{Ma:1995ey}, 
where $\vec{x}$ and $\vec{q}=q \hat{n}$ are comoving coordinates and momenta,
and $q$ and $\hat{n}$ are the momentum magnitude and direction.
Assuming that the gas is approximately homogeneous, 
the phase-space distribution can be divided into a homogeneous and isotropic background part $f_0(q)$, 
plus a small inhomogeneous perturbation $\Psi(\vec{x},\vec{q})$, 
\begin{equation}
f(\vec{x},\vec{q})=f_0(q)\big[1+\Psi(\vec{x},\vec{q})\big] \quad .
\label{eq:zerothorder}
\end{equation}
By definition, the zero-th order phase space distribution in Eq.~\eqref{eq:zerothorder} is required to be a solution of the Boltzmann equation in the homogeneous expanding background,
\begin{equation}
\frac{\partial f_0}{\partial \tau}
=
C[f_0] \quad , 
\label{eq:bgboltzmann}
\end{equation}
where $\tau$ is the conformal time, the collision terms on the right hand side can be read off from~\cite{Gorbunov:2011zzc}, and in the partial derivative the comoving momentum $\vec{q}$ and position $\vec{x}$ are held fixed.

In order to study the cosmological evolution and structure formation within SIDM, 
we need to analyze the evolution of both the homogeneous background distribution $f_0(q)$ and the small inhomogeneous perturbations $\Psi(\vec{x},\vec{q})$ defined in Eq.~\eqref{eq:zerothorder}.
We start by studying the background thermodynamics, 
and postpone the treatment of the perturbations to later sections. 

The main background quantities of the dark-matter fluid
are the homogeneous number density, energy density, and pressure.
They are obtained by taking moments of the distribution function $f_0$, 
and are given by  
\begin{eqnarray}
\nonumber n
&=&
\frac{1}{a^3}\int d^3q f_0(q) \quad ,
\\
\nonumber
\rho
&=&
\frac{1}{a^4}\int d^3q f_0(q) \epsilon(q) \quad ,\\
p
&=&
\frac{1}{3 a^4}\int d^3q \frac{q^2}{\epsilon(q)} f_0(q)  \quad ,
\label{eq:backgroundquantities}
\end{eqnarray}
where we defined the energy
\begin{equation}
\epsilon\equiv\sqrt{q^2+a^2 m^2} \quad. 
\label{eq:comovingenergy}
\end{equation}

From the background quantities the equation of state parameter $w$ is defined as the ratio of the pressure and energy density,
\begin{equation}
w \equiv {\frac{ p}{ \rho}} \quad .
\label{eq:eos}
\end{equation}
The speed of sound, on the other hand,
may be defined as the adiabatic propagation speed of small harmonic wave perturbations over the homogeneous background. 
This speed is given by
\begin{equation}
c_s=\frac{\dot{p}}{\dot{\rho}} \quad ,
\label{eq:speedofsound}
\end{equation}
where dots represent derivatives with respect to conformal time.
If dark matter is in kinetic equilibrium due to elastic self interactions, 
the common ansatz for the background phase space distribution is either the redshifted Fermi or Bose distributions~\cite{Gorbunov:2011zz}
\begin{equation}
f_0=\frac{g_s}{(2\pi)^3} \bigg[ {\exp\Big[\frac{\epsilon-a\mu}{aT(a)}\Big] \pm 1}\bigg]^{-1} \quad ,
\label{eq:equilibriumdistro}
\end{equation}
where the plus and minus signs correspond to fermions and bosons, respectively, $T$ is the gas temperature, and $g_s$ the parameter setting the degrees of freedom of the dark-matter particle. 
At early times, when dark matter is relativistic, Eq.~\eqref{eq:equilibriumdistro} reduces to 
\begin{equation}
f_0^{\textrm{rel}}=\frac{g_s}{(2\pi)^3}\bigg[ \exp\Big(\frac{q-\mu_0^R}{T_0^R}\Big) \pm 1 \bigg]^{-1}\quad ,
\label{eq:equilibriumdistrorel}  
\end{equation}
where we replaced the physical (time-dependent) temperature and chemical potentials $T(a)$ and $\mu(a)$ by their comoving values $T_0^R$ and $\mu_0^R$, 
using the usual temperature and chemical potential redshift expressions in the relativistic regime (which are obtained from particle number and entropy conservation, see e.g.~\cite{Kolb:1990vq}),
\begin{eqnarray}
\begin{array}{ccc}
T(a) &=& T_0^R/{a}  \quad ,  \\
\mu(a)
&=&
 {\mu_0^R}/{a}  \quad .
\end{array}
\label{eq:Tr}
\end{eqnarray}
Since $T_0^R$ and $\mu_0^R$ are quantities that are set at chemical decoupling, \textit{i.e.}, they are fixed by the dark-matter production mechanism, we refer to them as ``primordial.''

As the universe expands and dark matter cools down, the phase-space distribution may evolve in different ways. 
First, if dark matter kinetically decouples deep in the relativistic regime (at temperatures $T\gg m$), as in models of warm dark matter, the phase-space distribution remains frozen in the fully relativistic form Eq. \eqref{eq:equilibriumdistrorel}. 
In this case,  
the present-time ($a=1$) number and energy densities of  dark matter are obtained by using Eq.~\eqref{eq:equilibriumdistrorel} in Eq.~\eqref{eq:backgroundquantities}, and are given by
\begin{eqnarray}
\nonumber n&=&\mp \,g_s \frac{(T_0^R) ^3}{\pi^2} \textrm{Li}_3(\mp e^{\mu_0^R/T_0^R}) \\ 
 \rho&=&mn\left(1+  6 \left(\frac{T_0^R}{m}\right)^2 \frac{\textrm{Li}_5(\mp e^{\mu_0^R/T_0^R})}{ \textrm{Li}_3(\mp e^{\mu_0^R/T_0^R})}\right)\quad ,
 \label{eq:densityrel}
\end{eqnarray}
where $\textrm{Li}_n$ is the polylogarithm function of order $n$, and 
the minus and plus signs in their arguments correspond to fermions and bosons, respectively.
The second term on the right hand-side in the last line corresponds to the present-time kinetic energy,
which was obtained by using the fact that dark matter must be non-relativistic today, so we may approximate $\epsilon(a=1)=\sqrt{q^2+m^2}\sim m+q^2/2m$ in the integrand of Eq.~\eqref{eq:backgroundquantities}.

A second possibility is that dark matter remains kinetically coupled until it is non-relativistic, which is possible in models of SIDM. 
In this case, at $T\ll m$ 
the distribution function Eq.~\eqref{eq:equilibriumdistro} reduces to the Maxwell-Boltzmann function, 
\begin{equation}
f_0^{\textrm{n.r.}}=\frac{g_s}{(2\pi)^3}\exp\Big[\frac{\mu_0-m}{T_0}\Big]  \exp\Big[-\frac{q^2 }{2m T_0}\Big]  \quad ,
\label{eq:nrequilibriumdistro} 
\end{equation}
where we replaced the physical temperatures and chemical potentials by their comoving values at present times $T_0,\mu_0$ using 
\begin{eqnarray}
\begin{array}{cc}
 T(a) =  T_0  a^{-2}  \quad ,\\
\mu(a)=m+(\mu_0-m)\,{T(a)}/{T_0}  \quad  \quad \quad  \quad .
\end{array} 
\label{eq:Tnr}
\end{eqnarray}
In this case, 
the present time ($a=1$) number and energy densities of dark matter are obtained by using Eq.~\eqref{eq:nrequilibriumdistro} in Eq.~\eqref{eq:backgroundquantities}, and are given by
\begin{eqnarray}
\nonumber n&=&g_s \bigg(\frac{T_0 m}{2\pi}\bigg)^{3/2}  \exp\Big[\frac{\mu_0-m}{T_0}\Big]\\ 
 \rho&=&mn\left(1+ \frac{3 T_0}{2m} \right)\,, 
  \label{eq:densitynonrel}
\end{eqnarray}
where as before, we expanded the energy density using $\epsilon(a=1)=\sqrt{q^2+m^2}\sim m+ q^2/2m$. Expression Eq.~\eqref{eq:densitynonrel} is valid both for fermions and bosons.

A final possibility is that dark-matter decouples while semi-relativistic, at temperatures $T\sim m$. 
In the sudden decoupling approximation, the distribution function is in this case obtained by fixing the scale factor and temperatures at decoupling in Eq. \eqref{eq:equilibriumdistro} \footnote{A discussion on corrections to the sudden decoupling approximation can be found in \cite{Bernstein:1988bw,Trautner:2016ias}.}.
An even simpler alternative is to use the sudden transition approximation, 
where we  approximate the frozen-out distribution functions by Eq. \eqref{eq:equilibriumdistrorel} if the particle decouples while relativistic, and to Eq. \eqref{eq:nrequilibriumdistro} if it decouples while non-relativistic. 
For our purposes, \textit{i.e.} studying the observable effects of self interactions in the matter-power spectrum, using the sudden transition approximation will suffice, for two reasons.
First, these effects are phenomenologically most relevant if the dark sector is kinetically coupled when modes that matter for the \Lya forest or other small-scale observables enter the horizon. 
We will show that this requires large cross section values that also ensure that the dark sector is coupled until deep into the non-relativistic regime, in which case its distribution function is to an excellent approximation Boltzmann-like. 
Second, even when studying the phenomenological signatures left by models with smaller cross sections, we will show in concrete examples that the precise form of the dark-matter phase-space distribution has a numerically small effect on observables associated with the small-scale matter-power spectrum.
Thus, in what follows we will commit to the sudden transition approximation. 
We discuss further details of this approximation and the process of kinetic decoupling in sections \ref{sec:phasespace} and \ref{sec:decoupling}.

Now, it is important to note that the present-time temperature and chemical potential $T_0,\mu_0$ in the distribution Eq. \eqref{eq:nrequilibriumdistro}, which corresponds to a particle that decouples while non-relativistic,
are different from their primordial comoving values $T_0^R,\mu_0^R$.
In other words, a dark sector that when relativistic had a temperature and chemical potential given by Eq.~\eqref{eq:Tr} and that decouples while relativistic, has a present-time temperature $T_0^R$ and chemical potential $\mu_0^R$. 
On the other hand, a dark sector that starts with the same temperature and chemical potential as in Eq.~\eqref{eq:Tr}, but stays kinetically coupled as it becomes non-relativistic, evolves in such a way that it has a present temperature and chemical potential $T_0,\mu_0\neq T_0^R,\mu_0^R$.
With that being said, $T_0,\mu_0$ are not independent parameters, as they can be related to the primordial temperatures $T_0^R,\mu_0^R$ by computing the dark-matter evolution.
These relations are important for our purposes, as in order to fairly compare bounds on WDM versus bounds on SIDM, we must compare models that differ only by their self-interaction strengths, but that otherwise have the same primordial temperatures and chemical potentials.\footnote{More precisely and as we will see later, for a given particle statistics, this is equivalent to comparing models that have the same comoving or present-time velocity dispersion.} 

The relations between $T_0,\mu_0$ and $T_0^R,\mu_0^R$ can be found in two ways. 
The first one is to calculate the evolution of kinetically coupled dark matter as it goes through the semi-relativistic regime around $T\sim m$.
Around these temperatures, the phase-space distribution evolves semi-adiabatically from its
 relativistic Fermi/Bose form into the non-relativistic Boltzmann function, 
and the temperature and chemical potential have a non-trivial evolution, which differs from the simple expressions Eqs.~\eqref{eq:Tr} or \eqref{eq:Tnr}, but which matches onto Eqs.~\eqref{eq:Tr} and \eqref{eq:Tnr} at high and low temperatures correspondingly~\cite{Bernstein:1988bw}.
An alternative and much simpler way to obtain $T_0,\mu_0$ in terms of $T_0^R,\mu_0^R$ is by relating them using the fact that dark-matter number and energy densities are not modified by elastic collisions, as we now discuss.


\subsection{Matching the relativistic and non-relativistic regimes in the presence of self interactions}

As noted above, the presence of collisions leads to a non-trivial evolution of the dark-matter temperature and chemical potential in the relativistic to non-relativistic transition period. 
However, if the collisions are elastic they cannot change the number and kinetic energy densities that dark matter ends up having at present times. 
As a consequence,
these two quantities can be calculated at present times either in the presence or absence of collisions, 
resulting in Eq.~\eqref{eq:densityrel} or Eq.~\eqref{eq:densitynonrel},
and both results must coincide.
This leads to two relations between $T_0^R,\mu_0^R$ and $T_0,\mu_0$, 
given by 
\begin{eqnarray}
\nonumber \mp g_s\,\frac{(T_0^R) ^3}{\pi^2} \textrm{Li}_3(\mp e^{\mu_0^R/T_0^R})&=&  g_s \bigg(\frac{T_0 m}{2\pi}\bigg)^{3/2}  \exp\Big[\frac{\mu_0-m}{T_0}\Big] \quad \quad {\textrm{(number density matching)}}
\\
 \label{eq:matching0}
\\
\nonumber
\mp 6g_s \,\frac{(T_0^R)^5}{m\pi^2} \textrm{Li}_5(\mp e^{\mu_0^R/T_0^R})&=& \frac{3g_s}{4 \sqrt{2} \pi^{3/2}} T_0^{5/2} m^{3/2} \exp\Big[\frac{\mu_0-m}{T_0}\Big]  \quad \quad {\textrm{(energy density matching)}}
\\
\label{eq:matching}
\end{eqnarray}
where the minus and plus signs in their arguments corresponds to fermions and bosons, respectively.

The matching conditions can be simplified by equating the first line of Eq.~\eqref{eq:densitynonrel} to the present dark-matter number density $\rhoDM/m$, $\rhoDM=1.26 \times 10^{-6} \, \textrm{GeV}/\cm^3$ \cite{Aghanim:2018eyx}. 
In this way, 
we may express the present-time chemical potential $\mu_0$ as function of the dark-matter density,
\begin{equation}
\exp\Big[\frac{\mu_0-m}{T_0}\Big]  = \bigg(\frac{2 \pi}{T_0 m}\bigg)^{3/2} \frac{\rhoDM}{g_s m} \quad ,
\label{eq:chempot}
\end{equation}
so that using Eq.~\eqref{eq:chempot} in the matching conditions \eqref{eq:matching0} and \eqref{eq:matching}, 
we can rewrite these conditions in the simplified form 
\begin{eqnarray}
\mp\,\frac{g_s(T_0^R)^3}{\pi^2} \textrm{Li}_3(\mp e^{\mu_0^R/T_0^R})&=&  \frac{\rhoDM}{m} \quad \quad \quad {\textrm{(number density matching)}}
\label{eq:matching2}
\\
\mp 6 \,\frac{g_s(T_0^R)^5}{m\pi^2} \textrm{Li}_5(\mp e^{\mu_0^R/T_0^R})&=&\frac{3}{2} \frac{ T_0 \rhoDM }{m} \quad \quad {\textrm{(energy density matching)}}.
\label{eq:matching3}
\end{eqnarray}
Expressions \eqref{eq:matching2} and \eqref{eq:matching3} allow us to obtain the primordial temperature and chemical potentials in terms of the present-time dark-matter density and temperature, or vice-versa, 
as long as dark matter remains kinetically coupled through the transition from the relativistic to non-relativistic regimes.

All functions of conserved quantities 
are also unaffected by the elastic collisions. One particularly important such function is the dark-matter velocity dispersion, which by energy conservation stays constant as it is simply proportional to the kinetic energy.
We first define the comoving velocity $u$ by
\begin{equation}
u \equiv q/m \quad ,
\label{eq:veldef}
\end{equation}
and note that the comoving velocity $u$ coincides with the present-time particle velocity in the non-relativistic regime. 
Thus, the present-time velocity dispersion squared $\left<u^2\right>/3$ is given by
\begin{equation}
\frac{1}{3}\left<u^2\right>= \frac{1}{3\rhoDM/m}\int d^3q \frac{q^2}{m^2} f_0(q) \quad .
\end{equation}
For fermions and bosons presently described by the relativistic distribution function Eq.~\eqref{eq:equilibriumdistrorel}, the comoving velocity dispersion is 
\begin{equation}
\frac{1}{3}\left<u^2\right> = \mp \frac{4 g_s (T_0^R)^5}{m\pi^2 \rhoDM }  \textrm{Li}_5(\mp e^{\mu_0^R/T_0^R}) = 4\,\frac{ \textrm{Li}_5(\mp e^{\mu_0^R/T_0^R})}{ \textrm{Li}_3(\mp e^{\mu_0^R/T_0^R})} \bigg(\frac{T_0^R}{m}\bigg)^2  
\label{eq:velocitydispersionrel}
\end{equation}
where the minus and plus signs correspond to fermions and bosons respectively, and in the second equality we made use of Eq. \eqref{eq:matching2} to illustrate that $\left<u^2\right> \propto (T_0^R/m)^2$. 
On the other hand, if the dark matter remains kinetically coupled and is currently described by the non-relativistic distribution Eq.~\eqref{eq:nrequilibriumdistro}, its comoving velocity dispersion squared is
\begin{equation}
\frac{1}{3}\left<u^2\right> = \frac{T_0}{m} \quad .
\label{eq:velocitydispersionnr}
\end{equation}
Since the comoving velocity dispersion is not affected by collisions, given a primordial temperature $T_0^R$, 
$\usq$ is the same either if the dark matter decouples relativistically (leading to \eqref{eq:velocitydispersionrel}) or non-relativistically (leading to \eqref{eq:velocitydispersionnr}).
In fact, as a consistency check it is easy to see that equating \eqref{eq:velocitydispersionnr} and \eqref{eq:velocitydispersionrel} leads to the same matching condition Eq.~\eqref{eq:matching3} that fixes $T_0$ as a function of $T_0^R$.

A particular example that we will discuss later extensively, is the \textit{special case} of a vanishing primordial chemical potential, $\mu_0^R=0$.
This case is interesting as it is the closest to models of WDM, which chemically decouple while relativistic (and without large asymmetries in global quantum numbers) so their phase-space distribution is expected to have $\mu_0^R=0$. 
In this case, the matching conditions \eqref{eq:matching2} and \eqref{eq:matching3} reduce to 
\begin{eqnarray}
\frac{ g_s\xi_n \zeta(3)}{\pi^2} (T_0^R)^3&=&    \frac{\rhoDM}{m}  \label{eq:matching4}
\\
\frac{g_s\xi_\rho \zeta(5)}{m \pi^2} (T_0^R)^5&=& \frac{3}{2} \frac{ T_0 \rhoDM }{m}  ,
\label{eq:matching5}
\end{eqnarray}
where $\zeta$ is the Riemann zeta function, and $\xi_n,\xi_\rho$ are numerical coefficients. 
For fermions they are $\xi_n=3/4$, $\xi_\rho=45/8$, while for bosons $\xi_n=1$, $\xi_\rho=6$. 
Combining Eqns.~\eqref{eq:matching4} and \eqref{eq:matching5}, we obtain the relation
\begin{equation}
T_0= \frac{ 2\xi_\rho \zeta(5)}{3\xi_n \zeta(3)} \frac{(T_0^R)^2}{m}. 
\label{eq:temperaturerelation}
\end{equation}

In the special case $\mu_0^R=0$, the velocity dispersion of the dark matter if its current phase-space distribution is non-relativistic is simply Eq.~\eqref{eq:velocitydispersionnr}, while if it is relativistic it is given by
\begin{equation}
\frac{1}{3}\left<u^2\right> = \frac{4 g_s   \xi_\rho  \zeta(5) (T_0^R)^5}{6 m\pi^2 \rhoDM }=\frac{ 4 \xi_{\rho} \zeta(5)}{6 \xi_n \zeta(3)}  \bigg(\frac{T_0^R}{m}\bigg)^2 \label{eq:velocitydispersionrel2}
\end{equation}
where in the second equality we made use of \eqref{eq:matching4}.
Note that in order for relativistic bosonic or fermionic relics to have the same velocity dispersion today, their primordial temperatures $T_0^R$ must be slightly different, given the differences in $ \xi_v$.
Also, note that using Eq.~\eqref{eq:matching4}, one obtains that the velocity dispersion is $\sqrt{\usq/3} \propto T_0^R/m$, as expected.

Using Eq.~\eqref{eq:matching4}, we can also calculate the dark-matter mass required to obtain the correct relic abundance as a function of its primordial temperature. 
We obtain
\begin{equation}
m=
\frac{\pi^2}{\xi_n \zeta(3)}\frac{\rhoDM}{g_s (T_0^R)^3}
=
\left\{
\begin{array}{cc}
 4.1 \, \bigg[ \frac{0.1 \, \TSM }{T_0^R} \bigg]^3 \bigg[ \frac{2}{g_s} \bigg] \, \textrm{keV}  &  \quad \textrm{fermions}\\
 3.1 \, \bigg[ \frac{0.1 \, \TSM}{T_0^R} \bigg]^3 \bigg[ \frac{2}{g_s} \bigg] \, \textrm{keV}  &  \quad \textrm{bosons} \quad ,\\
\end{array}
\right.
\label{eq:wdmmass}
\end{equation}
where $T_{\rm SM}$ is the current CMB temperature.  
Alternatively, using Eqns.~\eqref{eq:velocitydispersionrel} and \eqref{eq:velocitydispersionnr}, the masses can be obtained in terms of the current velocity dispersion.
For dark matter that decouples while either relativistic or non-relativistic, we obtain
\begin{equation}
m=
\sqrt{\pi}
\bigg[\frac{\rhoDM}{g_s}\bigg]^{1/4}
\bigg[\frac{2\zeta(5)\xi_\rho }{\left<u^2\right>}\bigg]^{3/8}
\bigg[\frac{1}{\xi_n \zeta(3)}\bigg]^{5/8}
=
\left\{
\begin{array}{cc}
 4.7 \, \bigg[\frac{10^{-8}}{\sqrt{\left<u^2\right>/3}} \bigg]^{3/4} \bigg[ \frac{2}{g_s} \bigg]^{1/4} \, \textrm{keV}  &  \quad \textrm{fermions}\\
 4.0 \, \bigg[\frac{10^{-8}}{\sqrt{\left<u^2\right>/3}} \bigg]^{3/4} \bigg[ \frac{2}{g_s} \bigg]^{1/4} \, \textrm{keV}  &  \quad \textrm{bosons} \quad .\\
\end{array}
\right.
\label{eq:wdmmass2}
\end{equation}


\subsection{Phase-space distributions in the sudden transition approximation}
\label{sec:phasespace}
Throughout this work, the exact relations between the primordial and present temperatures and chemical potentials derived above will be used. 
However, the precise shape of the phase-space distribution during the transition from the relativistic to the non-relativistic regimes, 
which as commented before is less important, 
will be approximated by committing to a sharp transition between the distributions Eq.~\eqref{eq:equilibriumdistrorel} and \eqref{eq:nrequilibriumdistro} at a transition scale factor $\anr$.
In order to avoid discontinuities in the speed of sound, 
which will play an important role in the study of structure formation,
we choose the transition scale factor to be the one for which the speed of sound of a relativistic fluid $c_s^R=1/\sqrt{3}$
matches the speed of sound of a non-relativistic fluid $c_s^{\textrm{n.r.}}=\sqrt{\frac{\gamma T}{m}}$
(where $\gamma=5/3$ is the monoatomic polytropic index).
At this scale factor, the physical temperature of the dark matter is
\begin{equation}
T_{\textrm{NR}} \equiv \frac{m}{3\gamma}=\frac{1}{5} \, m  \quad .
\label{eq:tnrdef}
\end{equation}
The scale factor at the transition temperature can now be found using Eq.~\eqref{eq:Tnr}, giving
\begin{equation}
\anr=\sqrt{\frac{3\gamma T_0}{m}} = \sqrt{\gamma \left<u^2\right>} \quad ,
\label{eq:anr}
\end{equation}
where $\gamma=5/3$ and in the second equality we made use of Eq.~\eqref{eq:velocitydispersionnr}, to show that the transition scale factor $\anr$ is, up to an order one number, equal to the present dark-matter velocity dispersion $\sqrt{\usq/3}$.
In terms of $\anr$, the speed of sound in our approximation is simply given by
\begin{equation}
c_s=
\left\{
\begin{array}{cc}
1/\sqrt{3} & \quad a<\anr \\
\anr/(\sqrt{3}a) & \quad a>\anr  \quad ,
\end{array}
\right.
\label{eq:approxcs}
\end{equation}
which is a manifestly continuous function across the transition scale factor $\anr$.
We stress that our expression for the speed of sound is \textit{exact} deep in the relativistic and non-relativistic regimes, and is only approximate around $a\sim \anr$.

Summarizing, within our sudden transition approximation,
the self-interacting dark-matter distribution function is
\begin{equation}
f_0(u) =
\frac{g_s}{(2\pi)^3}\bigg[ \exp\Big(\frac{mu-\mu_0^R}{T_0^R}\Big) \pm 1 \bigg]^{-1} \quad \quad \quad \quad \,  a \leq \anr
\label{eq:equilibriumdistrorel2}
\end{equation}
\begin{equation}
f_0(u) =
 \frac{g_s}{(2\pi)^3}\exp\Big[\frac{\mu_0-m}{T_0}\Big]  \exp\Big[-\frac{m u^2 }{2 T_0}\Big] 
 \quad \quad \quad \,  a > \anr
\label{eq:nrequilibriumdistro2}
\end{equation}
where $\anr$ is given by Eq.~\eqref{eq:anr}, the primordial temperature and chemical potential $T_0^R$ and $\mu_0^R$ are exactly related to their present values  $T_0,\mu_0$ by Eq.~\eqref{eq:matching}, and in going from Eqns.~\eqref{eq:equilibriumdistrorel} and \eqref{eq:nrequilibriumdistro} to Eqns.~\eqref{eq:equilibriumdistrorel2} and \eqref{eq:nrequilibriumdistro2} we made use of Eq.~\eqref{eq:veldef}.

We conclude this section by commenting on a few important points regarding the background distribution functions Eqns.~\eqref{eq:equilibriumdistrorel2} and \eqref{eq:nrequilibriumdistro2}.
First, 
up to the order one dark-matter parameter $g_s$,
the distribution functions are completely described by three parameters:
the dark-matter mass $m$, 
and its temperature and chemical potential.
These three parameters are not independent, 
as a relation between them is set by fixing the dark-matter abundance via Eq.~\eqref{eq:matching2}.
In practice, we choose the primordial chemical potential $\mu_0^R$ and the comoving velocity dispersion $\sqrt{\usq/3}$ (or equivalently $\anr$, \textit{c.f.} Eq.~\eqref{eq:anr}) given in Eqn.~\eqref{eq:velocitydispersionrel}, as the two independent parameters describing the distribution functions.
We choose these parameters, since the primordial chemical potential can be directly related with the way in which dark matter was produced, and the velocity dispersion determines if dark matter is hot, warm, or cold.
In addition, and as discussed previously, the velocity dispersion has the convenient property that it is unaffected by collisions, \textit{i.e.}, 
dark matter particles with some given current velocity dispersion 
and a given spin, have the same primordial temperature $T_0^R/a$ at early times, 
regardless if the dark matter kinetically decoupled while relativistic or non-relativistic. 
Second, the distribution functions Eqns.~\eqref{eq:equilibriumdistrorel2} and \eqref{eq:nrequilibriumdistro2} 
are independent of the scale factor $a$.
This is as a result of the adiabatic evolution of the temperature and chemical potentials. 
As a consequence, 
from Eq.~\eqref{eq:backgroundquantities} we see that in the relativistic regime, 
where we set the energy to $\epsilon=q$, 
the dark-matter energy density redshifts as $1/a^4$,
while in the non-relativistic regime, with $\epsilon=am$,
the energy density redshifts as $1/a^3$, as expected.


\subsection{Kinetic decoupling}
\label{sec:decoupling}
As discussed above, if dark matter remains in kinetic equilibrium up to late times its background distribution function adiabatically evolves from the Fermi/Bose relativistic form 
into the Boltzman form, 
thanks to redistribution of momenta due collisions.
However, for small enough cross sections, dark matter may fall out of kinetic equilibrium while still relativistic, as in models of warm dark matter, or while semi-relativistic.
Kinetic equilibrium is lost once the self-interaction time $\tau(a)$ becomes larger than the Hubble time.
\begin{equation}
\tau(a)\equiv \bigg [ \frac{\rhoDM(a)}{m} \left< \sigma v(a,\anr) \right>\bigg]^{-1} \geq H^{-1}(a_d)\quad ,
\label{eq:taudef}
\end{equation}
where $a_d$ is the scale factor at decoupling and $v$ is the relative velocity between dark-matter particles.
As an example, for the special case of decoupling during radiation domination, 
using Eq.~\eqref{eq:taudef} and approximating the relative dark-matter velocity by $v=1$ for $a < \anr$ and  $v=\anr/a$ for $a>\anr$,
we obtain the decoupling scale factor,\footnote{Here we have taken the scattering cross section to be constant in order to obtain the decoupling redshift. This is a rather unrealistic assumption as in concrete particle physics models the cross section usually has some momentum dependence, especially in the relativistic regime. We will come back to this issue in section~\ref{sec:singlet}, where we study a concrete model realization of SIDM. }
\begin{equation}
a_\textrm{d}
=
5.67 \cdot 10^{-3} \, \sqrt{\anr} \bigg[\frac{\sigma/m}{10^{-5}~\textrm{cm}^2/\textrm{g}}\bigg]^{1/2} .
\label{eq:adnr}
\end{equation}

After decoupling,
the distribution function remains frozen in the relativistic form Eq.~\eqref{eq:equilibriumdistrorel2} if $a_d \leq \anr$ (as in warm dark-matter models),
in the non-relativistic form Eq.~\eqref{eq:nrequilibriumdistro2} if $a_d > \anr$ (as in typical WIMP models), 
or in a semi-relativistic form if $a_d \sim \anr$.
The subsequent evolution of the dark matter, 
in particular of the dark-matter perturbations to be studied in the following section, 
depends on the frozen-out distribution function.
Since we make use of the sudden transition approximation, 
we simply freeze out the dark-matter phase-space distribution in its fully relativistic form \eqref{eq:equilibriumdistrorel2} if $a_d \leq \anr$, 
or in the fully non-relativistic form \eqref{eq:nrequilibriumdistro2} if $a_d > \anr$. 

We comment now briefly on when errors may arise by committing to the sudden transition approximation.
If the dark-matter distribution freezes out at a scale factor $a_d\sim \anr$, 
it does so in a semi-relativistic shape, 
which as discussed before differs from the fully relativistic or non-relativistic forms.
This happens only for a specific combination of the two model parameters of self-interaction cross section and transition scale factor $\anr$ (or dark-matter velocity dispersion).
This coincidental combination is obtained by setting $a_d=\anr$ in Eq.~\eqref{eq:taudef},
and is given by the special value $a_{d,\textrm{NR}}$.
As an example, for the special case of decoupling during radiation domination this gives
\begin{equation}
a_{d,\textrm{NR}}
=
3.22 \cdot 10^{-5} \bigg[\frac{\sigma/m}{10^{-5}~\textrm{cm}^2/\textrm{g}}\bigg]  \quad \quad \quad  \quad .
\label{eq:anrcoincidence}
\end{equation}
As a consequence, we expect our sudden transition approximation to have somewhat larger errors in the regions of parameter space where \eqref{eq:anrcoincidence} is approximately fulfilled. 
In our analysis of the dark-matter perturbations we will clearly identify the regions of parameter space where the dark matter decouples semi-relativistically, in order to know where there might be small errors that arise from our approximation.



\section{Cosmological dark-matter perturbations}
\label{sec:cosmopert}

We now move on to the calculation of the evolution of  self-interacting dark-matter perturbations on top of the homogeneous background.

Dark-matter perturbations behave differently in the kinetically coupled and decoupled regimes.
In the former case, perturbations are described by oscillating sound waves, which imprint a dark sound horizon on the matter-power spectrum.
After decoupling, on the other hand, 
perturbations free-stream and are smoothed out due to transport from over-dense into under-dense regions.
The sound horizon and free-streaming scales are the most relevant dynamical quantities for understanding dark-matter perturbations,
so we dedicate section \ref{sec:soundhorizon} to study the parametrics of these two simple thermodynamic scales. 

A precise calculation of the matter-power spectrum
requires solving numerically the Boltzmann equations for the evolution of the dark matter and Standard Model perturbations.
With this purpose, we implement a new Boltzmann code that computes the evolution of the dark matter and Standard Model perturbations.\footnote{We found that implementing all the required changes in existing Boltzmann solvers such as CLASS~\cite{Lesgourgues:2011rh} would require significant modifications over the public versions, 
which was more cumbersome to us than implementing our own code.}
We divide the evolution of the dark matter into two regimes---kinetically coupled and decoupled---and 
describe the corresponding evolution equations in sections \ref{sec:evolution1} and \ref{sec:evolution2}.
Differently from the case of cold non-interacting dark matter, 
the evolution must account for the possibility of dark-matter diffusion, free streaming and sound waves, 
and for the change of a relativistic into a non-relativistic phase-space distribution.

\subsection{Free-streaming and sound waves}
\label{sec:soundhorizon}

In the absence of self interactions, 
a dark-matter particle free streams at its physical velocity $dx/d\eta=v(\eta)$, where $x$ are comoving coordinates and $\eta$ is the conformal time.
Free-streaming suppresses dark-matter perturbations on scales below the distance travelled by dark matter, 
which is given by
\begin{equation}
 \ell_{fs}(\eta)\equiv\int_0^{\eta}{d\eta'\, v(\eta')} \quad .
\label{eq:fsdef}
\end{equation}
Since growth of matter perturbations happens mostly during matter domination,
a useful quantity to evaluate the suppression of perturbations is the free-streaming length up to matter-radiation equality $ \ell_{fs}(\eta_{\textrm{eq}})$~\cite{Kolb:1990vq}.
This scale can be easily estimated by splitting the integral over the early period in which the dark matter travels relativistically, 
and over the later period in which its velocity redshifts with the scale factor as $1/a$.
The transition between the two periods happens at a scale factor $\anr$ defined in Eq.~\eqref{eq:anr}, 
which is (up to an order one factor) equal to the present dark-matter velocity dispersion.
Approximating $v=1$ for $a<\anr$ and $v=\anr/a$ for $a>\anr$ we obtain
\begin{eqnarray}
\nonumber \ell_{fs}(\anr,a)|_{a=\aeq}&=&
\int_{0}^{\anr} \frac{da}{H a^2}
+
\int_{\anr}^{\aeq} \frac{da}{H a^2} \frac{\anr}{a}
\\
\nonumber
&=&
\frac{1}{H_0 \sqrt{\aeq}}\, 
\bigg[
\int_{0}^{\anr} da
+
\int_{\anr}^{\aeq} da \frac{\anr}{a}
\bigg]
\\
&=&
\frac{\anr}{H_0 \sqrt{\aeq}}
\bigg[1+\ln\bigg(\frac{\aeq}{ \anr}\bigg)\bigg]
\label{eq:fs}
\end{eqnarray}
where in going to the second line we approximated Hubble at radiation domination by $H(a)=H_0 \sqrt{\aeq}/a^{2}$, 
and in each line the first and second terms correspond to free streaming during the relativistic and non-relativistic regimes. 
Note that both the relativistic and non-relativistic period end up contributing in a similar order to the free-streaming length, $\sim \anr/(H_0\sqrt{\aeq})$, 
with the non-relativistic piece having in addition an order one logarithmic enhancement on top. 
Numerically, we obtain
\begin{equation}
\ell_{fs}(\anr,a)|_{a=\aeq} 
=
233 \,\, \textrm{kpc}  \,\bigg[\frac{\anr}{10^{-7}}\bigg]\bigg[\frac{1+\ln\big(\aeq/\anr\big)}{1+\ln\big(\aeq/10^{-7}\big)}\bigg] \quad,
\label{eq:fsnumber}
\end{equation}
where we normalized to scales of order $\mathcal{O}(100 \textrm{kpc})$, since they are the smallest scales that can be probed using the Lyman-$\alpha$ forest, 
and the logarithmic enhancement factor is $\ln\big(\aeq/10^{-7}\big)\simeq 8$.
From Eq.~\eqref{eq:fsnumber} we see that free-streaming dark matter with velocity dispersion $\sim \anr \gtrsim 10^{-7}$ would be in tension with Lyman-$\alpha$ measurements (a much more precise analysis presented later will indicate that the Lyman-$\alpha$ bounds are closer to  $ \anr \gtrsim \textrm{few} \times 10^{-8}$).

If dark matter has self interactions, on the other hand, 
it does not free stream but there is still suppression of power at small scales, 
now due to pressure support. 
The relevant length scale to describe self-interacting dark matter is the sound horizon, 
defined as 
\begin{equation}
r_s(\eta)  \equiv\int_0^\eta d\eta' c_s(\eta')  \quad ,
\end{equation}
where $c_s$ is the dark-matter speed of sound Eq.~\eqref{eq:approxcs}. 
In particular, 
the sound horizon at the time of dark kinetic decoupling
gives an estimate of the characteristic scales below which the growth of perturbations is affected by pressure support. 
To simplify the estimation of the sound horizon, in this section we assume that dark matter decouples before matter-radiation equality (but otherwise in the rest of this work we allow for decoupling at all times).
With this assumption, the sound horizon at decoupling is given by 
\begin{eqnarray}
\nonumber
r_s(\anr,a)|_{a=a_d} 
&=&
\frac{\anr}{H_0 \sqrt{3\aeq}}
\bigg[1+\ln\bigg(\frac{a_d}{\anr}\bigg)\bigg]
\\
&\simeq&
\frac{233}{\sqrt{3}} \bigg[\frac{\anr}{10^{-7}}\bigg]\bigg[\frac{1+\ln\big(a_d/\anr\big)}{1+\ln\big(a_d/10^{-7}\big)}\bigg] \,\, \textrm{kpc}
\label{eq:soundhorizon}
\end{eqnarray}
where $a_d$ is the scale factor at decoupling, given as a function of the dark-matter cross section in Eq.~\eqref{eq:adnr}.
There are two important characteristics to point out regarding this result. 
First, note that the sound horizon Eq.~\eqref{eq:soundhorizon} is parametrically equal to the free-streaming scale  \eqref{eq:fs}, as they are both of order $ \anr/(H_0\sqrt{\aeq})$.
This means that the presence or absence of self interactions does not lead to a significant modification of the scale below which dark-matter perturbations are damped, as pointed out in~\cite{deLaix:1995vi}.
Qualitatively, however, free streaming and pressure support lead to different suppression mechanisms:
while free-streaming amounts to a monotonic decrease of power at small scales,
pressure support leads to acoustic oscillations. 
Second, 
note that the sound horizon depends only logarithmically on the self-interaction cross section, which enters in the decoupling scale factor $a_d$.
As a consequence, 
power suppression due to pressure support is mostly controlled by the velocity dispersion $\sqrt{\usq/3}\sim \anr$, 
and has a mild (but non-negligible) dependence on the particle physics parameters controlling the elastic self interactions. 

\begin{figure}[ht!]
\begin{center}
\includegraphics[width=8cm]{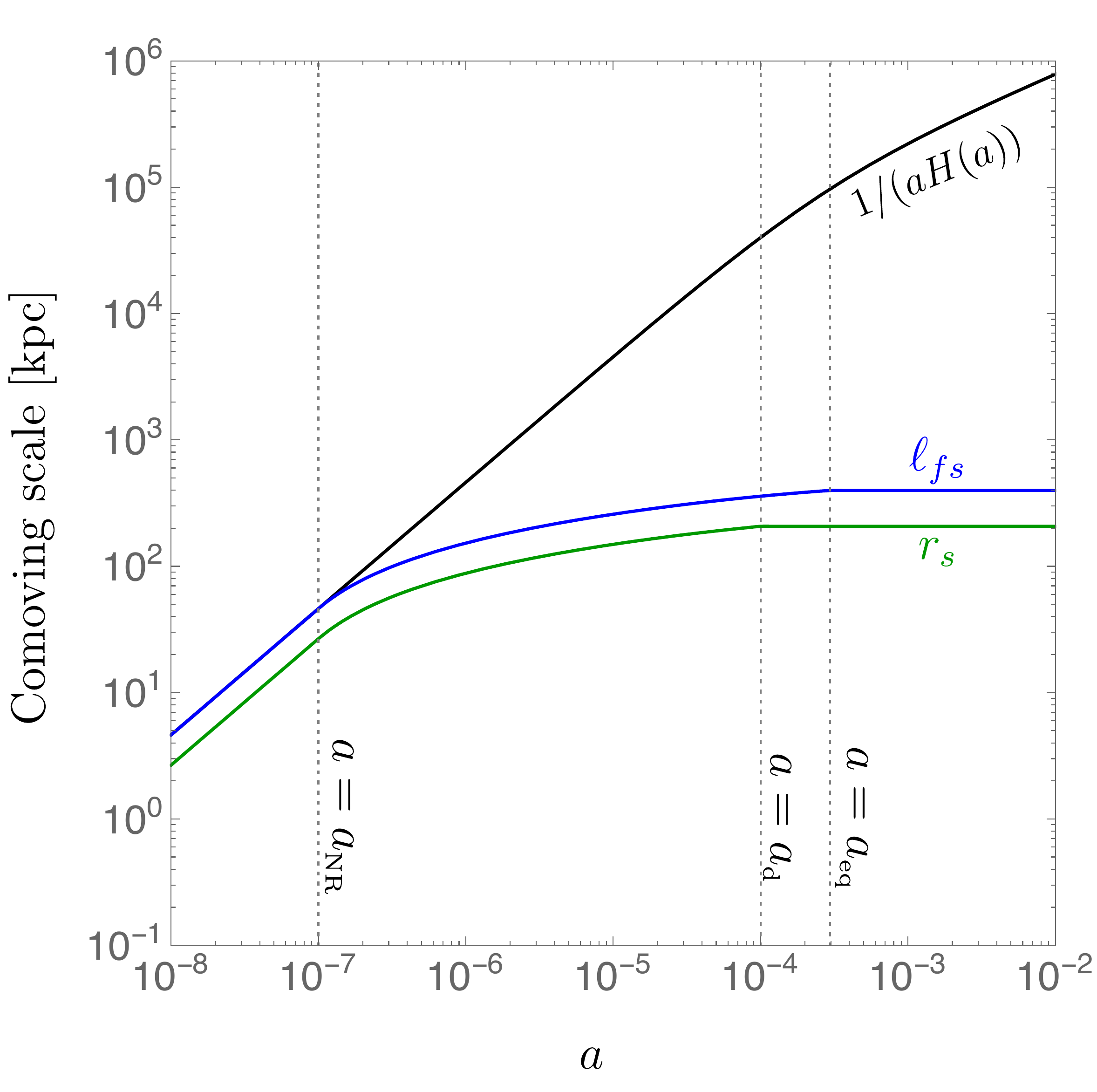}
\caption{Comoving sound-horizon $r_s$ and free-streaming $\ell_{fs}$ scales as a function of the scale factor, for a particle that self interacts or free streams. 
In black we also show the comoving horizon, $1/(aH(a))$. 
We assume that the self-interacting species kinetically decouples at $a_d=10^{-4}$, and becomes non-relativistic at $\anr=10^{-7}$. 
Up to decoupling, the difference between the sound horizon and free-streaming lengths is only a factor of $1/\sqrt{3}$.}
\label{fig:rs}
\end{center}
\end{figure}

We compare the sound horizon and free-streaming lengths in Fig.~\ref{fig:rs}, 
for decoupling and transition scale factors $a_d=10^{-4}$, $\anr=10^{-7}$.
At large redshift, both the free-streaming and sound horizon lengths grow linearly with the scale factor, and follow closely the comoving horizon size $1/aH(a)$,
but at $a\sim \anr$ they moderate their growth to logarithmic due to the redshift of the particle velocity or speed of sound. 
From the figure, we also clearly see that both the free-streaming scale and the sound horizon asymptote to values of the same order at present times.

\subsection{Evolution of dark-matter perturbations: kinetically coupled regime}
\label{sec:evolution1}

The dark-matter perturbations over the homogeneous background are contained in the function $\Psi$ defined in Eq.~\eqref{eq:zerothorder}.
Following the treatment and notation of~\cite{Ma:1995ey}, 
we expand the perturbation in a Legendre series,  
\begin{equation}
\Psi(\vec{k},q,\hat{n},a)=\sum_{l=0}^\infty (-i)^l (2l+1) \Psi_l(\vec{k},q,a) P_l(\hat{k}\cdot \hat{n}) \quad ,
\end{equation}
where $\vec{k}$ is the comoving spatial Fourier mode and $\vec{q}= q \hat{n}$ is the comoving particle momentum.

The first two moments of the Legendre hierarchy are related to the fluid's density $\deltaDM$ and velocity perturbations $\thetaDM$ by
\begin{eqnarray}
\nonumber
\deltaDM &=& \frac{4\pi}{\rhoDM a^4} \int q^2 dq \epsilon f_0(q) \Psi_0 \quad ,\\
\thetaDM &=& \frac{4\pi k}{\rhoDM(1+w) a^4} \int q^3 dq f_0(q) \Psi_1  \quad ,
\label{eq:densityandvelocity}
\end{eqnarray}
where $\epsilon$ is the energy Eq.~\eqref{eq:comovingenergy} and  $w$ is the equation of state parameter Eq.~\eqref{eq:eos}.
The second moment is related to the fluid's anisotropic stress $\sigmaDM$
\begin{eqnarray}
\sigmaDM &=& \frac{8\pi}{3\rhoDM (1+w)a^4} \int q^2 dq \frac{q^2}{\epsilon} f_0(q) \Psi_2  \quad .
\label{eq:viscosity}
\end{eqnarray}
(The fluid's anisotropic stress should not be confused with the dark-matter self-interaction cross section, for which the same letter $\sigma$ is conventionally used.) Working in the conformal Newtonian gauge, 
the evolution equations for the dark-matter density and velocity perturbations are~\cite{Ma:1995ey}
\begin{eqnarray}
\nonumber 
\dot{\delta}_{\textrm{DM}}&=&
-(1+w)(\thetaDM-3\dot{\phi})
-3 H_\eta \big[ c_s^2-w\big] \deltaDM \quad ,
\\
\dot{\theta}_{\textrm{DM}}&=&
-H_\eta
(1-3w)\thetaDM
-
\frac{\dot{w}}{1+w}\thetaDM
+
\frac{c_s^2}{1+w}
k^2 \deltaDM
-k^2  \sigmaDM
+k^2\psi \quad ,
\label{eq:fluidperts}
\end{eqnarray}
where all derivates are with respect to conformal time, 
and the conformal Hubble parameter is defined as
\begin{equation}
H_\eta\equiv\frac{\dot{a}}{a} \quad .
\end{equation}
In Eq.~\eqref{eq:fluidperts} $c_s$ is the adiabatic dark-matter speed of sound.
The variables $\phi$ and $\psi$ are the conformal Newtonian potential perturbations. 
Their evolution is given by
\begin{eqnarray}
\nonumber
\dot{\phi}&=&
\frac{1}{3 H_\eta}
\Big[
-4\pi G a^2
\Big(
\rho_{\gamma}(a) \delta_{\gamma}
+
\rho_{\nu}(a) \delta_{\nu}
+
\rhoDM(a) \delta_{\textrm{DM}}
+
\rho_{b}(a) \delta_{b}
\Big)
-3 H_\eta^2 \psi
-k^2 \phi
\Big] \quad ,
\\
\psi
&=&
-\frac{1}{k^2}
\big[
12\pi G a^2 (\bar{\rho}+\bar{P}) \sigma
\big]
+\phi \quad ,
\label{eq:potentials}
\end{eqnarray}
where the indices $\gamma,\nu,b$ refer to the Standard Model photon, neutrino, and baryons, 
and $ (\bar{\rho}+\bar{P}) \sigma $ is the anisotropic stress weighted and averaged over all the matter and radiation components according to
\begin{equation}
 (\bar{\rho}+\bar{P}) \sigma \equiv \sum_i (\rho_i + P_i) \sigma_i \quad ,
 \label{eq:viscositytotal}
\end{equation}
with $i=\textrm{DM}, \gamma,\nu,b$,
and $\rho_i$ and $P_i$ are the background energy density and pressure of each component.
We describe the treatment of the Standard Model perturbations in appendix \ref{app:SM}.

For a scale factor $a<a_d$ (c.f.~Eq.~\eqref{eq:adnr}) dark matter is kinetically coupled and 
can be treated as a close to perfect fluid.
Collisions ensure that higher moments of the Boltzmann distribution, $\Psi_{l\geq 2}$ remain small, 
so that deviations from the perfect fluid situation can be characterized by small anisotropic stress terms, 
which are due to dark-matter diffusion. 
In this case, we make a series of approximations that significantly simplify the computation of the evolution of dark-matter perturbations.

First, the speed of sound $c_s$ can be approximated by expression Eq.~\eqref{eq:approxcs},
which correctly describes the speed of sound of kinetically coupled dark matter in the non-relativistic or relativistic regimes.
In the semi-relativistic regime such an expression leads to small errors in the evolution of perturbations. 
However, it is easy to see that such errors are not important for the phenomenological study of this model. 
As pointed out in section~\ref{sec:soundhorizon}, 
dark-matter models that are semi-relativistic at a scale factor equal or greater than $\anr \sim \textrm{few} \times 10^{-8}$ will be in tension with Lyman-$\alpha$ observations. 
From Fig.~\ref{fig:rs} we see that at such transition scale factor, 
the modes that are entering the horizon are of order $\sim 10$ kpc.
These scales are below what is observable with current power-spectrum probes, 
so the small errors of our approximation during the semi-relativistic regime are phenomenologically irrelevant.

Second, the equation of state parameter can be computed from its evolution equation given by~\cite{Hu:1998kj}
\begin{equation}
\dot{w}=3H_\eta (1+w)(w-c_s^2) \quad .
\label{eq:wparameter}
\end{equation}
Using Eq.~\eqref{eq:approxcs}, we approximate the solution of \eqref{eq:wparameter} to $w=c_s^2=1/3$ in the relativistic regime, and $w  \simeq T/m =3/5\,c_s^2$ in the non-relativistic regime. 

Finally, when the fluid is close to perfect the anisotropic stress can be computed  as a function of the  dark-matter velocity perturbation $\thetaDM$, without the need to compute higher moments in the Boltzmann hierarchy $\Psi_{l\geq 2}$, 
using the \textit{relaxation time approximation}.
We lay out the assumptions of this approximation and compute the resulting anisotropic stress in appendix~\ref{app:viscosity}. 
The relaxation time approximation gives
\begin{equation}
\sigmaDM=
\left\{
\begin{array}{cc}
\frac{4}{15 a} \thetaDM \tau(a) & \quad a< \anr \\
\frac{4 a}{15 \anr^2}  \thetaDM \tau(\anr)& \quad a > \anr
\end{array}
\right. \quad ,
\label{eq:viscosityapprox}
\end{equation}
where $\tau$ is the self-interaction time, given in Eq.~\eqref{eq:taudef}.
As expected, in the infinite-cross section limit, $\tau=0$, and the anisotropic stress vanishes so that dark matter reduces to a perfect fluid. 
Note also that within our quasi-perfect fluid approximation, 
the dark-matter perturbations can be evolved without any reference to the specific shape of the background dark-matter distribution $f_0$. 
Only averaged thermodynamic properties such as the speed of sound and equation of state parameter are relevant for calculating dark-matter perturbations while it remains kinetically coupled.

\subsection{Evolution of dark-matter perturbations: decoupled regime}
\label{sec:evolution2}
At the scale factor $a=a_d$ (\textit{c.f.} Eq.~\eqref{eq:adnr}) dark matter kinetically decouples and starts free streaming. 
The subsequent evolution cannot be computed within the close to perfect fluid approximation of the previous section,
as higher moments of the Boltzmann distribution become relevant and the relaxation time approximation breaks down. 
Therefore, for $a\geq a_d$ we directly compute the evolution of the higher moments as follows. 
First, 
for evolving modes (modes  inside the horizon) at $a=a_d$  we must match the dark-matter density and velocity perturbations $\deltaDM$ and $\thetaDM$, 
computed in the kinetically coupled regime, 
to the Boltzmann moments $\Psi_l$.
We calculate the matching conditions in appendix \ref{app:matching}. 
They are given by 
\begin{eqnarray}
\nonumber
\Psi_0 &=&
- \xi_\delta \frac{\deltaDM}{4}  \frac{d\ln f_0}{d \ln q} \\
\nonumber
\Psi_1 &=&
-\, \frac{\thetaDM}{3} \frac{\epsilon}{kq}  \frac{d\ln f_0}{d \ln q} \\
\nonumber
\Psi_2 &=&- \frac{2 \thetaDM   H_{\eta}^{-1}}{15 }   \frac{d\ln f_0}{d \ln q}  
\\
\Psi_{l\geq 3} &=& 0
\label{eq:matchingad}
\end{eqnarray}
where $\xi_\delta=1, \epsilon=q$ or $\xi_\delta=4/3, \epsilon=a_d m$ if the dark matter is relativistic or non-relativistic at matching. 
For non-evolving modes that are outside the horizon at $a=a_d$, 
the density and velocity perturbations are frozen before horizon entry.
In this case, 
we perform the matching at $a>a_d$ but much before horizon crossing, 
using $\xi_\delta=1, \epsilon=q$ or $\xi_\delta=4/3, \epsilon=a_d m$ if the dark matter is relativistic or non-relativistic at crossing. 

The subsequent evolution of the Boltzmann moments is then computed using~\cite{Ma:1995ey},
\begin{eqnarray}
\nonumber \dot{\Psi}_0 &=& \frac{-qk}{\epsilon} - \dot{\phi} \frac{d\ln f_0}{d \ln q} \\
\nonumber \dot{\Psi}_1 &=& \frac{qk}{3\epsilon}(\Psi_0-2\Psi_2) - \frac{\epsilon k}{3q} \psi \frac{d\ln f_0}{d \ln q} \\
 \dot{\Psi}_{l\geq 2} &=& \frac{qk}{(2l+1)\epsilon}\big [l \Psi_{l-1}-(l+1)\Psi_{l+1} \big]  \quad ,
 \label{eq:moments}
\end{eqnarray}
where $\phi$ and $\psi$ are the conformal Newtonian potentials, 
which are evolved according to Eq.~\eqref{eq:potentials}.
The dark-matter density, velocity, and anisotropic stress perturbations, 
required to solve for the potentials, 
are obtained by integrating the Boltzmann moments using Eqns.~\eqref{eq:densityandvelocity} and \eqref{eq:viscosity}.
Note that both the matching conditions Eq.~\eqref{eq:matchingad} and the evolution equations Eq.~\eqref{eq:moments} for the moments depend on the background phase space distribution $f_0$, 
which is frozen out in a relativistic or non-relativistic form depending when kinetic decoupling happens, 
as discussed in section \ref{sec:decoupling}.
The distribution is thus given by Eq.~\eqref{eq:equilibriumdistrorel2} for $a_d\leq\anr$ or by Eq.~\eqref{eq:nrequilibriumdistro2} for $a_d>\anr$.

Since the hierarchy of moments is infinite, it must be truncated. 
We use the truncation prescription of~\cite{Ma:1995ey},
\begin{equation}
\Psi_{l_{\textrm{max}}+1}=
\frac{(2l_{\textrm{max}}+1)\epsilon}{qk\eta} \Psi_{l_{\textrm{max}}}-\Psi_{l_{\textrm{max}}-1}
\end{equation}
The truncation must be done at a sufficiently high value of $l_{\textrm{max}}$ to avoid large errors. 
In our computations we take $l_{\textrm{max}}=30$, which we find to be sufficient.

\subsection{Numerical procedure}

To solve for the dark matter and Standard Model perturbations, 
we write a C program containing the code for the evolution equations. 
The numerical procedure starts at a redshift $z=10^{10}$, 
where modes of comoving size $\sim$0.05~kpc enter the horizon, 
with horizon-crossing initial conditions as in~\cite{Ma:1995ey,CAMB} 
and cosmological parameters from~\cite{Aghanim:2018eyx}. 
We solve the differential equations implementing the adaptive step size Runge-Kutta-Fehlberg RF45 algorithm. 
We decrease the algorithm's error tolerance until we find stable solutions. 
When solving for the dark-matter Boltzmann moments, 
we discretize the velocity space $q= m u$ (c.f.~Eq.~\eqref{eq:veldef}) in 30 steps from $u=0$ to $u= 6\anr$. 
We checked that increasing $q$ beyond the chosen upper limit or decreasing the discretization size of $q$ space does not lead to significant changes in our results. 
We perform the $q$ integrals in Eqns.~\eqref{eq:densityandvelocity} and \eqref{eq:viscosity} using Riemann's method.

\section{Results}
\label{sec:results}

We now present the results of our analysis of the power spectrum of self-interacting dark matter.
In section \ref{sec:ps}, we present the power spectrum for different choices of the self-interaction cross section and dark-matter velocity dispersion. 
In section \ref{sec:lymanalpha}, we set limits using Lyman-$\alpha$ observations. 

\subsection{SIDM power spectrum}
\label{sec:ps}

The self-interacting dark matter-power spectrum is controlled mostly by two parameters: the current dark-matter velocity dispersion $\sqrt{\usq/3}$ and the cross section over the dark-matter mass $\sigma/m$.
In addition, in order to specify the dark-matter phase-space distribution, 
the primordial dark-matter chemical potential $\mu_0^R$ needs to be fixed.
In this section for simplicity,
and to be able to directly compare with warm dark-matter models, 
we set $\mu_0^R=0$ (we comment on the $\mu_0^R\neq 0$ case in the next section).
Note that for $\mu_0^R=0$, the dark-matter mass is fixed by its velocity dispersion using Eqns.~\eqref{eq:wdmmass} and \eqref{eq:matching5} when dark matter kinetically decouples while non-relativistic.

\begin{figure}[t!]
\begin{center}
\includegraphics[width=5cm]{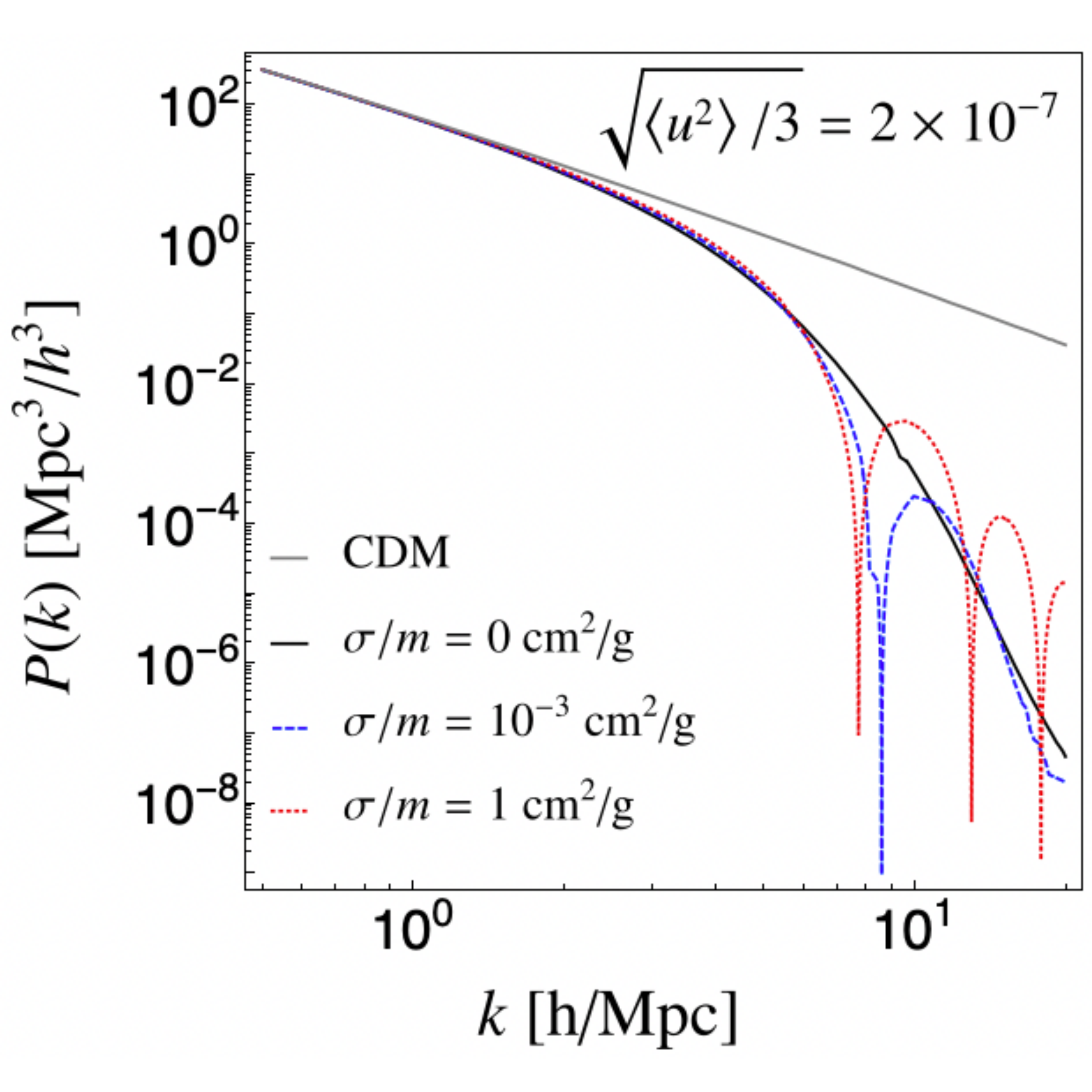}
\includegraphics[width=5cm]{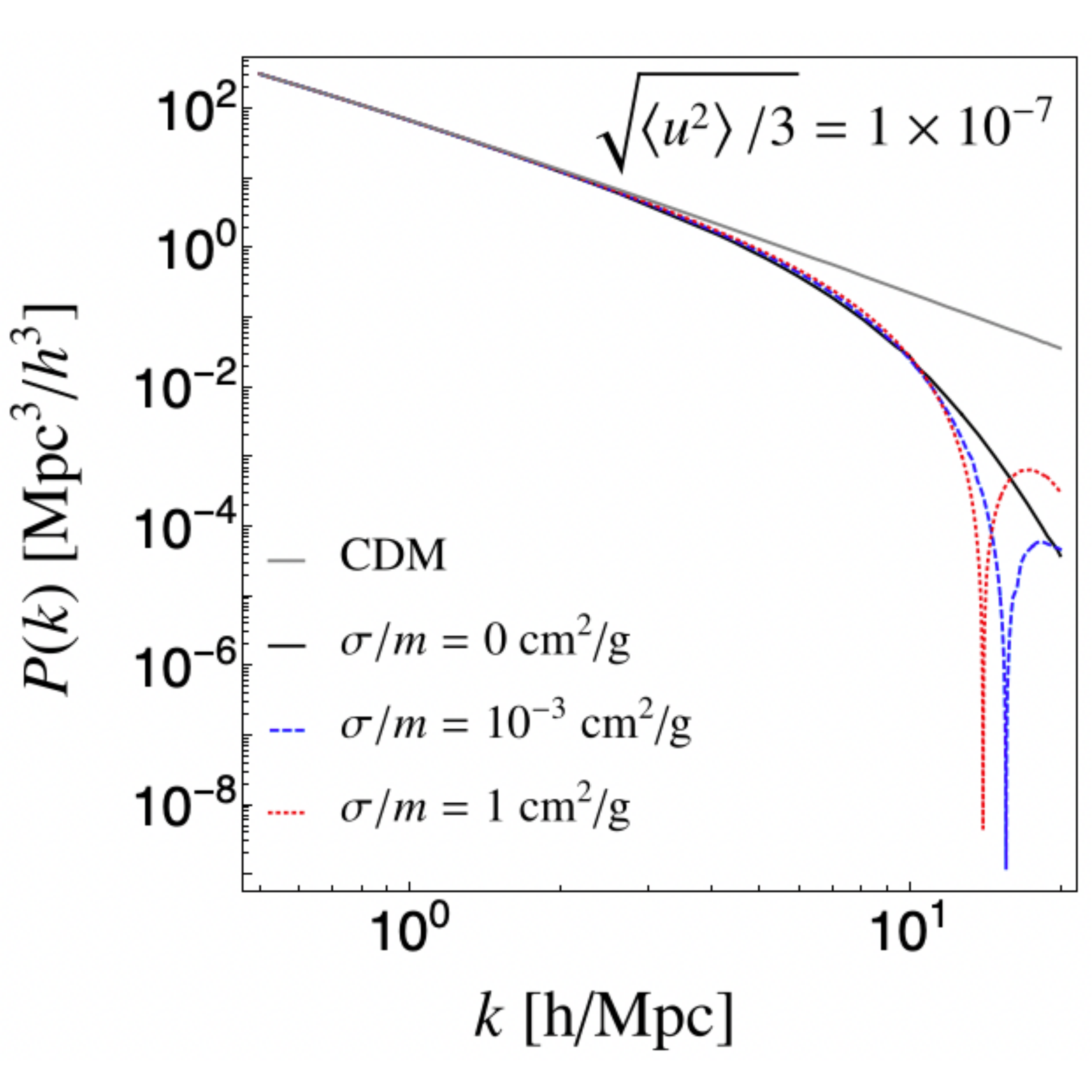}
\includegraphics[width=5cm]{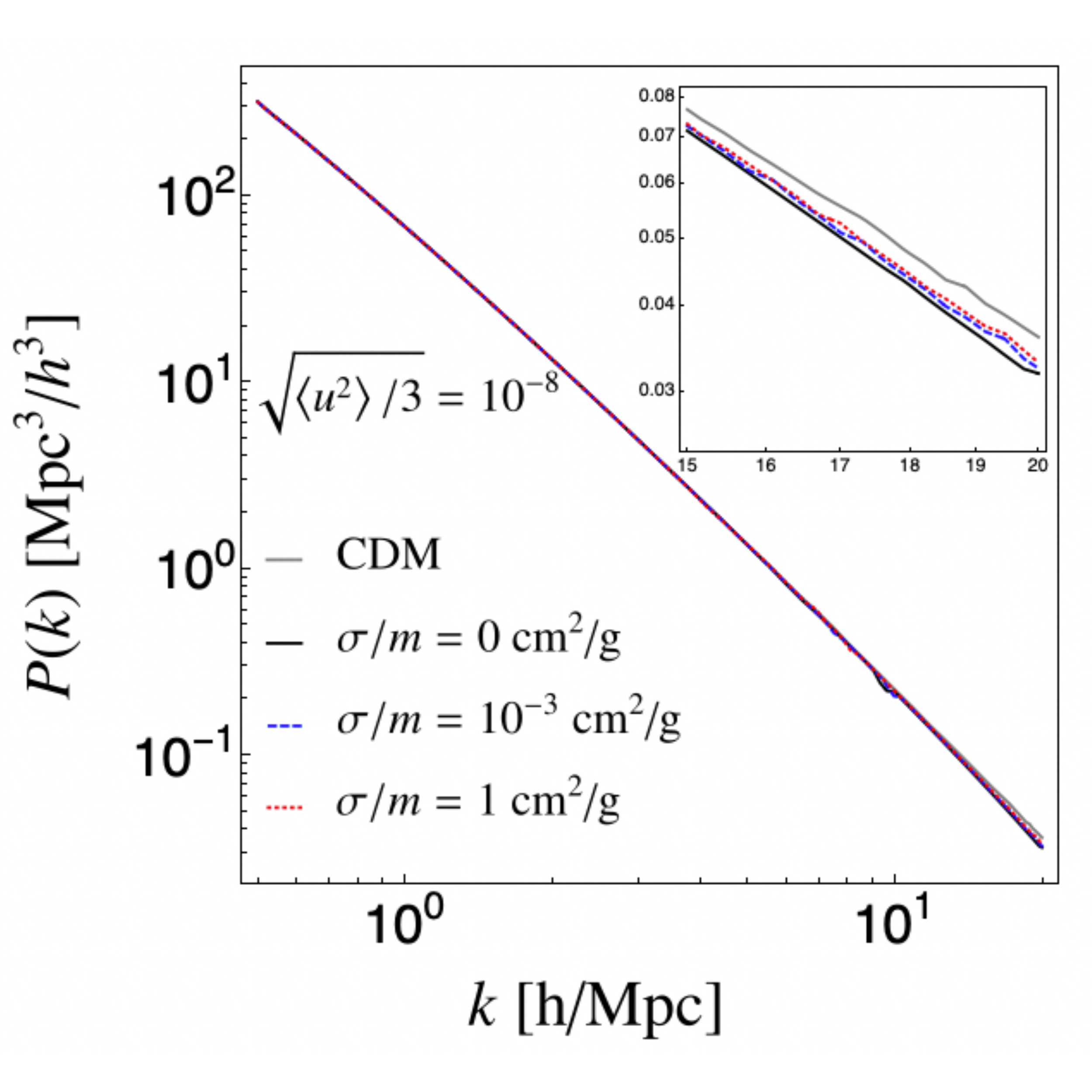}
\caption{Linear matter-power spectrum for SIDM with velocity dispersion $\sqrt{\usq/3}=2\times 10^{-7}$ (left panel), $\sqrt{\usq/3}=10^{-7}$ (middle panel), and $\sqrt{\usq/3}=10^{-8}$ (right panel), for elastic self-interaction cross sections $\sigma/m=0$, $\sigma/m=10^{-3}\cm^2/\g$ and $\sigma/m=1\cm^2/\g$.  For these velocity dispersions and for a fermion with a vanishing primordial chemical potential, $\mu_0^R=0$, the corresponding dark-matter masses are 0.49~keV, 0.83~keV, and 4.67~keV, respectively (see Eq.~\eqref{eq:wdmmass2}).}
\label{fig:PS1}
\end{center}
\end{figure}

With this choice, we present the dark-matter-power spectrum for $\sqrt{\usq/3}=2\times 10^{-7}$  (left panel), $\sqrt{\usq/3}=1 \times 10^{-7}$ (middle panel), and $\sqrt{\usq/3}=2 \times 10^{-8}$ (right panel), for three values of cross section in Fig.~\ref{fig:PS1}.
Let us first discuss the case $\sqrt{\usq/3}=2\times 10^{-7}$. 
Here we clearly see that for all three choices of cross section the power spectrum is suppressed at scales $k \gtrsim 5 \, \textrm{h/Mpc}$.
This is consistent with our discussion in section \ref{sec:soundhorizon}, 
where we found that for both self-interacting and collisionless dark matter, 
the suppression of power on small scales is determined mostly by the dark-matter velocity dispersion and has only a weak dependence on the cross section.
However, and as expected, the nature of the power suppression is different for the different values of cross section: 
for $\sigma/m=0$ we see a sharp and smooth power cutoff, which is due to free-streaming, 
while for $\sigma/m=1\, \cm^2/\g$ we see instead suppression of power due to acoustic oscillations, 
with the period of the momentum-space acoustic oscillations being determined by the inverse of the dark-matter sound horizon.
The situation for $\sigma/m= 10^{-3}\, \cm^2/\g$ is somewhat in-between the $\sigma/m=0$ and $\sigma/m=1\, \cm^2/\g$ cases. 
For $\sigma/m= 10^{-3}\, \cm^2/\g$ we see that the period of the momentum-space oscillations is a factor of few larger than for $\sigma/m=1\, \cm^2/\g$,
a feature that is indicative of the smaller sound-horizon obtained for smaller cross-sections, 
c.f.~Eq.~\eqref{eq:soundhorizon}.
From the figure it is clear that by measuring the period, or more generally, the full shape of the acoustic oscillations, information regarding the dark-matter self-scattering cross section can in principle be obtained.\footnote{
It is important to note however that current bounds on the SIDM velocity dispersion (to be discussed in section \ref{sec:lymanalpha}) are at the level of $\sqrt{\usq/3}\lesssim 1\times 10^{-8}$. 
Thus, the power spectrum of models that have acoustic oscillations and are not yet ruled out looks like the one presented in Fig. \ref{fig:PS1}, right panel. 
From that figure, it is clear that in allowed models the acoustic oscillations are fully visible only at very small scales, which are experimentally inaccessible at present.
}
In addition, from the figures we observe that models with self-interactions contain more power than WDM models at wave numbers   below the characteristic sound-horizon or free-streaming scales.
For instance, for  $\sqrt{\usq/3}=2\times 10^{-7}$, for which $\pi r_s^{-1}/2 \sim 10 \, \Mpc/h$, with $r_s$ being the sound horizon at matter-radiation equality, models with $\sigma/m=1\, \cm^2/\g$ have more power at $k\sim 3 h/\Mpc$ than $\sigma/m=0$ models with the same velocity dispersion.
This is a manifestation of the strength of the power suppression of WDM, which scales as a high power of wave number (as can be seen from analytic approximations to the WDM transfer function, see \textit{e.g.} \cite{Viel:2005qj}), 
while the envelope of pressure-induced power suppression characteristic of SIDM is relatively milder \cite{Kolb:1990vq}.
This feature is also partially associated to the fact that the free-streaming scale of WDM models is a factor of $\sqrt{3}$ larger than the sound-horizon scale of SIDM models, for the same velocity dispersion.

To illustrate the dependence of the power spectrum suppression on the velocity dispersion, take now the case $\sqrt{\usq/3}=1\times 10^{-7}$ presented in the right panel of Fig.~\ref{fig:PS1}. In this case, we see that the cutoff in the power spectrum moves towards higher values of wavenumber when compared with our previous larger choice of $\sqrt{\usq/3}$.
This result confirms the linear dependence of both the free-streaming length and sound-horizon scales on the velocity dispersion, discussed in section \ref{sec:soundhorizon}. 
For $\sqrt{\usq/3}=2\times 10^{-8}$ we still see suppression of power on small scales for all the three values of cross sections (we zoom on the the smallest scales in the inset for clarity). 
In this case, however, a whole acoustic oscillation is not completed up to scales $k=20\,h/\Mpc$ for the cases of non-vanishing $\sigma/m$.
Thus, up to these scales, the difference in self-interaction strengths can only be distinguished by the amount of power suppression, 
which still depends on $\sigma/m$, as clearly seen in the figure inset.

\subsection{Bounds from Lyman-$\alpha$}
\label{sec:lymanalpha}

The suppression of power on small scales due to acoustic oscillations or free-streaming leads to effects that can be measured in the flux power spectrum of Quasi Stellar Objects (QSO).
The flux power spectrum of such objects presents a series of hydrogen absorption peaks due to baryonic clouds along the line of sight to the QSO, 
which absorb light at the Lyman-$\alpha$ frequency, at different redshifts.
A decrease of the matter-power spectrum on small scales leads to fewer baryonic gas clouds, which in turn leads to less Lyman-$\alpha$ absorption.

In order to set bounds on dark-matter models using such probes, 
hydrodynamical simulations are required in order to obtain the flux-power spectrum for a given model and compare the results to data (see \textit{e.g.}~\cite{Irsic:2017ixq}). 
In order to circumvent these complications, 
here we make use of a simplified prescription developed in~\cite{Schneider:2016uqi,Murgia:2017lwo} referred to as the ``area criterion.''
This prescription allows us to set bounds on models by comparing their small-scale power spectrum suppression, quantified by an ``area estimator,'' with the corresponding suppression in a reference model known to be excluded from full  hydrodynamical simulations and data analyses.
The area estimator is calculated using the linear power spectrum
so a non-linear analysis is not required in order to set bounds with this prescription.
All the information regarding the dark-matter model being analyzed, SIDM in our case, is contained in the area estimator, 
while the  information regarding Lyman-$\alpha$ data is encoded in the reference model that sets the exclusion boundary.
In what follows we take the reference model to be fermionic $g_s=2$ warm dark matter with a mass $m=5.3 \, \keV$ (or $m=3.5 \, \keV$ for conservative \Lya bounds),
which sets the $2\sigma$ exclusion boundary according to an analysis of the MIKE, HIRES, and XQ-100 spectrometers in~\cite{Irsic:2017ixq}.
We discuss further details and the validity of the area criterion in appendix~\ref{app:areacriterion}, 
and we now move on to the analysis of the resulting bounds. 

\begin{figure}[t!]
\begin{center}
\includegraphics[width=7.5cm]{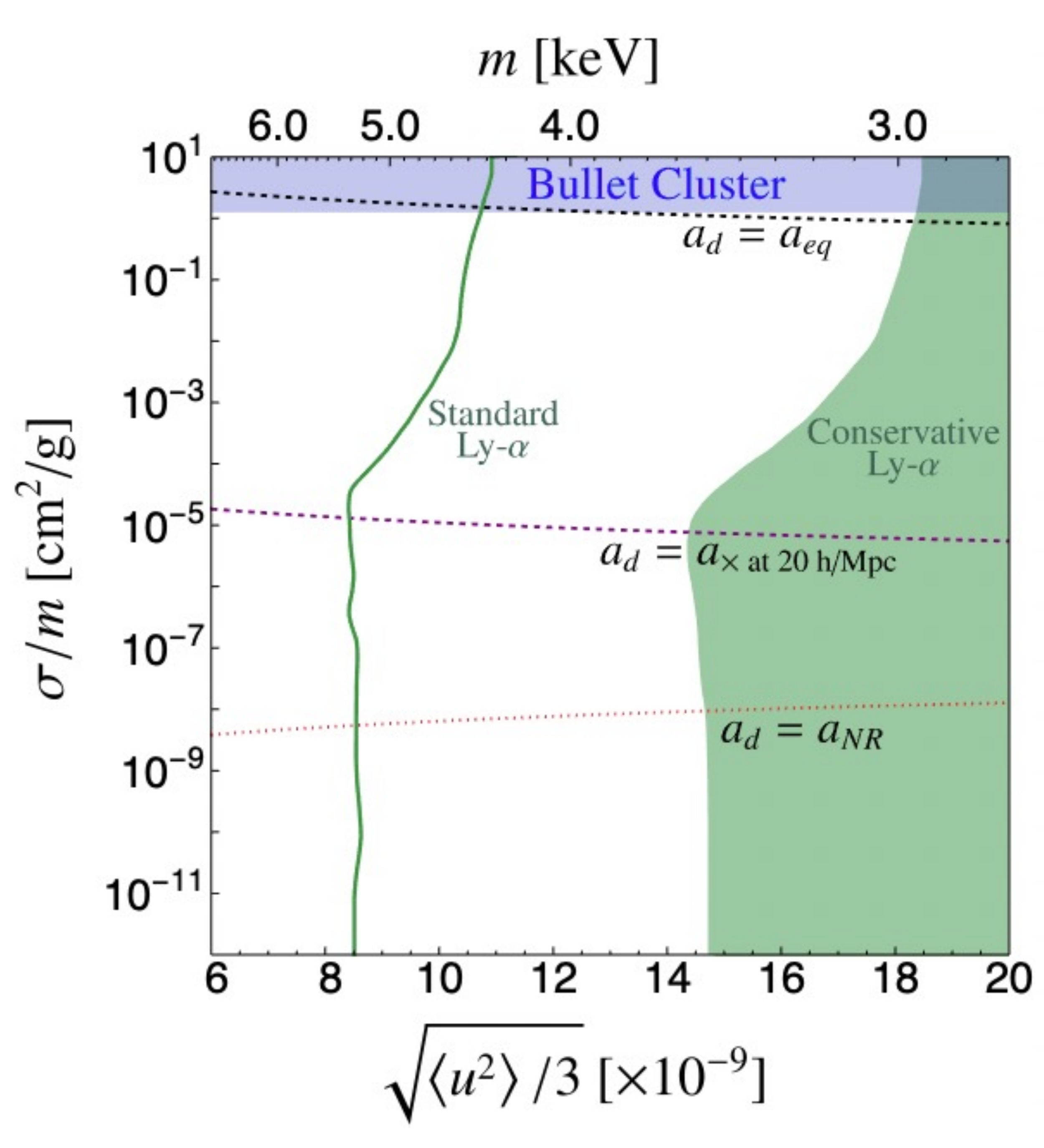}
\includegraphics[width=7.5cm]{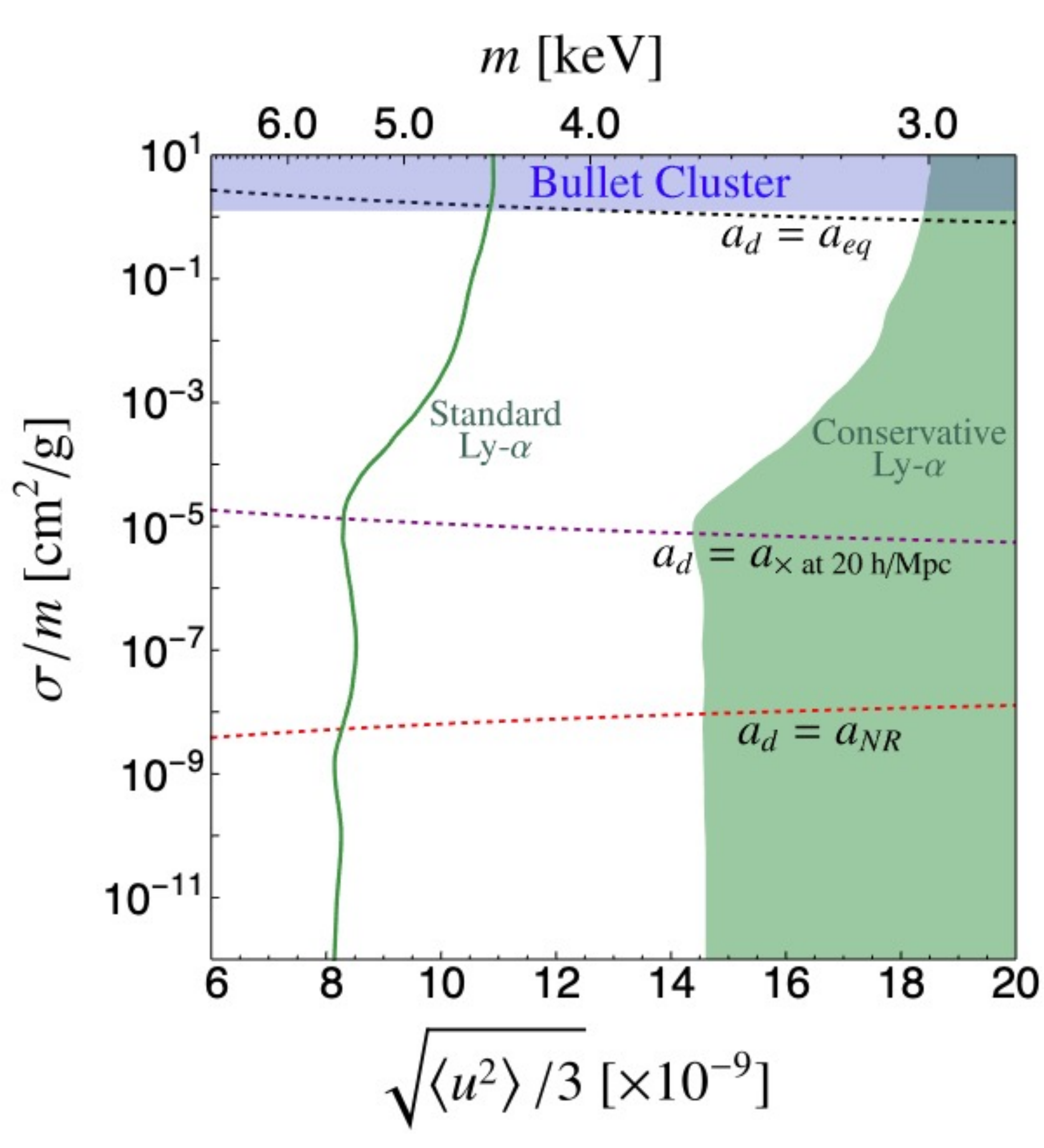}
\caption{Bounds from the \Lya forest at $95\%$ CL (green) and the Bullet Cluster at  $68\%$ CL \cite{Randall:2007ph} (blue) on fermionic (left panel) and bosonic (right panel) SIDM, 
as a function of the present-time velocity dispersion $\sqrt{\usq/3}$ and self-scattering cross section over mass, $\sigma/m$. 
In both cases we have set the primordial chemical potential to zero. 
For fermionic dark matter, we take a number of degrees of freedom $g_s=2$, corresponding to a single Weyl fermion, while for bosonic dark matter we take $g_s=1$, corresponding to single real scalar particle. 
To account for uncertainties in \Lya bounds, we follow~\cite{Irsic:2017ixq} and present a conservative limit in the shaded green region (that excludes standard fermionic WDM with $m\geq 3.5\,\keV$) and a standard limit with the solid green line (that excludes standard fermionic WDM with $m\geq 5.3\keV$).
The dashed purple line shows the cross section values below which dark matter has kinetically decoupled before \Lya modes have entered the horizon, which occurs at $k\sim 20 \textrm{h}/\Mpc$ and $a_{\times\ \textrm{at}\ 20\textrm{h/Mpc}}\simeq 10^{-6}$. 
The dotted red line $a_d=\anr$ shows the cross sections below which dark matter kinetically decouples while relativistic. 
Below this line the model corresponds to warm-dark matter.
}
\label{fig:fermiexclusion}
\end{center}
\end{figure}

The Lyman-$\alpha$ bounds on SIDM are presented in Fig.~\ref{fig:fermiexclusion} as a function of the dark-matter velocity dispersion $\sqrt{\usq/3}$ (or its mass, via Eq.~\eqref{eq:wdmmass}) and the self-scattering cross section, 
for the case of a fermion (left panel) or boson (right panel). 
In both cases, the primordial chemical potential has been set to zero, and we assumed that the dark matter can only be in kinetic equilibrium, but is out of chemical equilibrium. 
We show two bounds to account for \Lya uncertainties.
These uncertainties are due to the fact that different assumptions can be done regarding the temperature evolution of baryons (see \textit{e.g.}~\cite{Garzilli:2015iwa,Hui:2016ltb}), which affect \Lya absorption. 
Under two assumptions for such evolution, \cite{Irsic:2017ixq} finds two bounds on the mass WDM, a standard and a conservative one, which then translate into our two different bounds for SIDM using the area criterion. 
We show standard bounds with a solid green line, and in the filled green region we show conservative limits.
From the figures, we immediately see that  Lyman-$\alpha$ bounds have a mild but clear dependence on the particle's cross section, 
and they are weakest when the cross section is large. 
For the largest cross sections consistent with Bullet Cluster bounds, 
Lyman-$\alpha$ excludes fermionic $g_s=2$ (bosonic $g_s=1$) SIDM with a mass $m\leq 4.4 \, \keV$ ($m \leq 4.45\, \keV$). 
For a vanishing cross section, on the other hand, the bound reduces to that of warm dark matter (as it must, given our procedure for calculating the bound), which is $m\leq 5.3\,\keV$ for $g_s=2$ fermions. 

From the figures we also see that for all cross sections smaller than $\sigma/m \leq 10^{-5} \,\cm^2/\textrm{g}$, the bounds are essentially equal to the ones obtained for the $\sigma/m=0$  case.
This can be easily understood by comparing the scale factor at which the smallest mode relevant for the \Lya  forest $k\sim 20 \textrm{h}/\Mpc$ enters the horizon at $a_{\times\ \textrm{at}\ 20\textrm{h/Mpc}}\simeq 10^{-6}$, with the scale factor of kinetic decoupling $a_d$, given in Eq.~\eqref{eq:adnr}. 
These two scale factors are equal precisely around $\sigma/m \sim 10^{-5} \cm^2/\textrm{g}$, as shown by the dashed line in the figure. 
For cross sections smaller than this threshold value,
\Lya modes enter the horizon when dark matter has already kinetically decoupled and is free-streaming, so the bounds essentially reduce to those of WDM.
For larger cross sections,
\Lya modes enter the horizon while dark matter is acoustically oscillating.
In this case the power suppression is, on average, less than for a free-streaming species for the scales that can be currently probed (although there are clearly some ranges of scales on which acoustic oscillations lead to more suppression), 
so bounds get comparatively weaker.

In Fig.~\ref{fig:fermiexclusion} we also show with a dotted line the threshold cross section above which dark matter decouples when it already became non-relativistic, $a_d\geq \anr$ (c.f.~Eq.~\eqref{eq:anrcoincidence}).
From the line we see that for all cross sections that lead to observable acoustic oscillations in the \Lya forest (the parameter space above the line $a_d=a_{\times~\textrm{at}~20~\textrm{h/Mpc}}$), 
the cross section is large enough that the dark matter decouples while already non-relativistic.\footnote{As a matter of fact, 
since \Lya modes enter the horizon at $a_{\times \textrm{at} 20\textrm{h/Mpc}}\simeq 10^{-6}$, and the bounds presented in the figures impose that dark matter becomes non-relativistic at $\anr \sim \sqrt{\usq/3}\lesssim 10^{-8}$,
when \Lya modes enter the horizon dark matter must always be deep in the non-relativistic regime. }
This means that for the purposes of studying elastic particle interactions in acoustic oscillations,
dark matter can be treated as non-relativistic, 
and in particular, 
the errors discussed in section~\ref{sec:equilibriumdistro} from approximating the background distribution, speed of sound, and  equation of state as fully non-relativistic can be safely neglected. 
This comment also applies to other small-scale observables such as Milky Way satellite counts and strong lensing, as they currently test similar scales and velocity dispersions to the observed with the \Lya forest~\cite{Enzi:2020ieg}.

More generally, the exact shape of the background distribution function is not particularly relevant for calculating small-scale bounds, at least for the typical non-degenerate thermal distributions considered here.
The reason is that the free-streaming scale and sound-horizon that set the cutoff in power, Eqns.~\eqref{eq:fs} and~\eqref{eq:soundhorizon}, are mostly set by the particle's velocity dispersion, 
and do not depend much on the precise shape of the phase-space distribution. 
This has two consequences for \Lya bounds. 
First, it means that the bounds on the velocity dispersion on fermions and bosons are very similar, 
as can be seen by comparing the left and right panels in Fig.~\ref{fig:fermiexclusion}.
The only significant difference between the fermionic and bosonic cases is that for a given velocity dispersion and primordial chemical potential (set here to zero), the particle's mass required for the species to be $100\%$ of dark matter is different, as seen from Eq.~\eqref{eq:wdmmass}. 
As a consequence, the map between the lower $\sqrt{\usq/3}$ and upper $m$ axes in the figures slightly differ for fermions and bosons, and also when considering other values of the degree-of-freedom parameter $g_s$.
And second, 
the approximate independence from the background distribution of the bounds  
implies that it is not particularly important if the dark matter decouples while relativistic with a Fermi or Bose distribution, while non-relativistic with a Boltzmann distribution, or while semi-relativistic with a transitional distribution.
This can be seen explicitly by comparing the bounds below the $a_d=\anr$ line in Fig.~\ref{fig:fermiexclusion} and those between the $a_d=\anr$ and $a_d=a_{\times \textrm{at} 20\textrm{h/Mpc}}$. 
In these two regions observable \Lya modes enter the horizon when dark matter free streams with a relativistic and non-relativistic distribution, correspondingly. 
We see that in both cases bounds are almost equal, confirming their approximate independence on the underlying background distribution. 

\section{A benchmark example: singlet-scalar dark matter} 
\label{sec:singlet}
We now present a specific dark-matter model that satisfies the requirements of our general discussion so far. 
We take dark matter to be a real scalar field $S$, whose Lagrangian is
\begin{equation}
\mathcal{L_S} = \frac{1}{2} \partial_\mu S \partial^\mu S -\frac{m_S^2}{2} S^2 - \frac{\lambda}{4!} S^4\,.
\label{eq:Lphi}
\end{equation}
In Eq.~\eqref{eq:Lphi} we have imposed a $\mathbb{Z}_2$ symmetry under which $S$ is odd, 
that forbids cubic terms and super-renormalizable $SH^\dagger H$ couplings to the Higgs, so that the field remains stable. For vacuum stability, we also require $\lambda > 0$. We refer the reader to~\cite{Burgess:2000yq,Egana-Ugrinovic:2015vgy} for further details on this model.

In this work we have studied only velocity-independent self interactions, 
and have calculated the kinetic decoupling redshift and anisotropic stress Eq.~\eqref{eq:viscosityapprox} under this assumption.
In the scalar model,
the self-interaction cross section in the non-relativistic limit $T \ll m_S$ is indeed approximately constant and given by
\begin{equation}
\sigma(SS\rightarrow SS) = \frac{\lambda^2}{128 \pi m_S^2}\,.
\label{eq:sigmadec}
\end{equation}
In the relativistic limit, the cross section becomes momentum-dependent
and a semi-relativistic analysis of decoupling must be carried out. 
For simplicity, here we limit ourselves to the parameter space where the particle decouples non-relativistically, and use the constant cross section Eq.~\eqref{eq:sigmadec} throughout. 
Note that since \Lya modes enter the horizon when dark matter is deep in the non-relativistic regime, 
the anisotropic stress can always be calculated using the non-relativisic cross section Eq.~\eqref{eq:sigmadec}.

The singlet sector can be populated in the early universe by a variety of mechanisms, the most popular one being including a renormalizable quartic coupling to the Higgs, $ S^2 H^\dagger H$~\cite{Silveira:1985rk,mcdonald1994gauge,Burgess:2000yq}.
However, no significant couplings to the Standard Model sector are required for the singlet to be the dark matter, 
as it can be produced for instance through its couplings to the inflaton.
For details on singlet production from the inflaton we refer the reader to~\cite{Adshead:2016xxj,Arcadi:2019oxh}, 
and in what follows we simply parametrize the dark sector by its primordial chemical potential $\mu_R^0$ and current velocity dispersion $\sqrt{\usq/3}$, 
which set the number density of the dark sector.
Again, we do not consider any interactions with the SM.

As in previous sections we limit ourselves to a dark sector that is out of chemical equilibrium (but can be in kinetic equilibrium), so that number-changing interactions (singlet cannibalization~\cite{Carlson:1992fn,Pappadopulo:2016pkp}) alter
neither the relic abundance that we have assumed to be set primordially, potentially depleting the dark sector, nor the standard temperature redshift described in section~\ref{sec:equilibriumdistro}.
We take the minimal values of the quartic coupling $\lambda$ required for a real scalar to achieve chemical equilibrium with itself at any point of its thermal history from~\cite{Arcadi:2019oxh},  assuming a vanishing primordial chemical potential $\mu_0^R = 0$ for concreteness.

\begin{figure}[t!]
\begin{center}
\includegraphics[width=8cm]{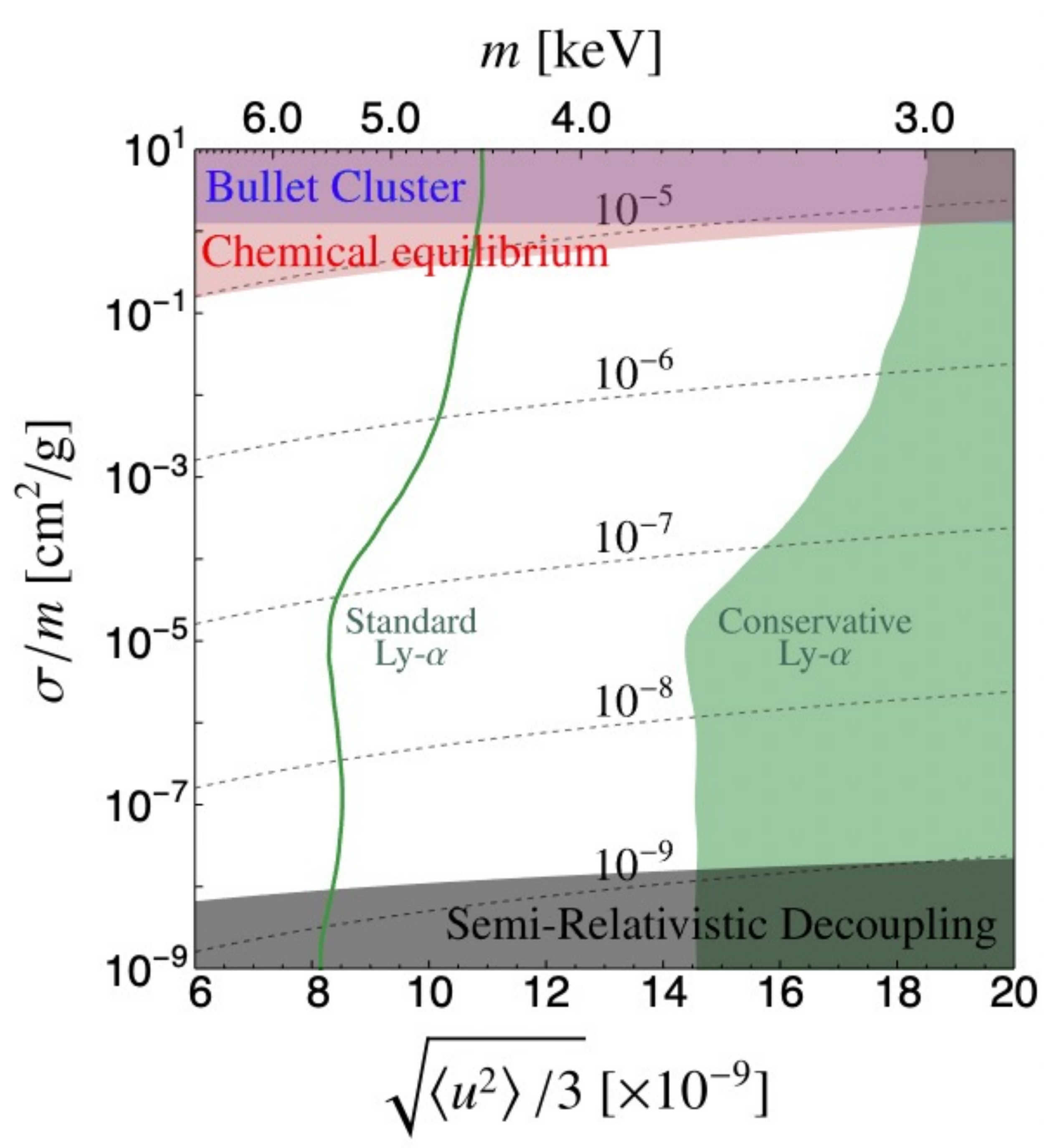}
\caption{Our exclusion (in green) for the scalar model, 
as a function of the present velocity dispersion $\sqrt{\usq/3}$ and self-scattering cross section over mass. 
In dashed lines we show values of the quartic self-coupling $\lambda$.
We also show existing bounds on the self-interaction cross section from the Bullet Cluster \cite{Randall:2007ph} (blue), as well as regions in which our conclusions do not apply either because the singlet may decouple relativistically (gray) or because it achieves chemical self equilibrium (red); see text for details. 
This plot was made for a spin-0 ($g_f = 1$) boson with no initial (primordial) chemical potential ($\mu_0^R = 0$). 
However, changing the $g_f$ factor of the particle or the primordial chemical potential changes mostly the conversion from $\sqrt{\usq/3}$ in the lower axis to the dark-matter mass $m$ in the upper axis required to obtain the correct relic abundance,
otherwise bounds on $\sqrt{\usq/3}$ remain similar.}
\label{fig:scalar}
\end{center}
\end{figure}

With these assumptions, we show our bounds as applied to the scalar-singlet model in Fig.~\ref{fig:scalar}. 
In the figure we have included in dashed-black the values of the quartic coupling corresponding to the given cross section according to Eq.~\eqref{eq:sigmadec}.
We also show in shaded red the regions where the scalar achieves chemical equilibrium and our analysis may become invalid, assuming $\mu_0^R=0$. 
The only important dependence on the primordial chemical potential $\mu_0^R$ in Fig.~\ref{fig:scalar} is in the calculation of the regions where the scalar reaches chemical equilibrium and in the map between the velocity dispersion in the lower axis to the particle's mass in the upper axis. 
Otherwise, our bounds on the velocity dispersion apply to all values of $\mu_0^R$ to a good approximation, since the free-streaming lengths and sound-horizon are mostly set by $\sqrt{\usq/3}$. 
In the figure we also show in shaded gray the values of cross sections below which a semi-relativistic analysis of kinetic decoupling must be carried out. 
For self couplings in the range $5\times 10^{-8}\lesssim \lambda \lesssim 10^{-5}$, where bounds on the scalar's velocity dispersion or mass weaken, \Lya modes enter the horizon when the scalar is kinetically coupled and acoustically oscillates, so bounds are weaker than for a free-streaming species. 
For couplings below $\lambda\lesssim 5\times 10^{-8}$ the scalar mostly free streams and behaves as warm dark matter instead.

\begin{figure}[t!]
\begin{center}
\includegraphics[width=8cm]{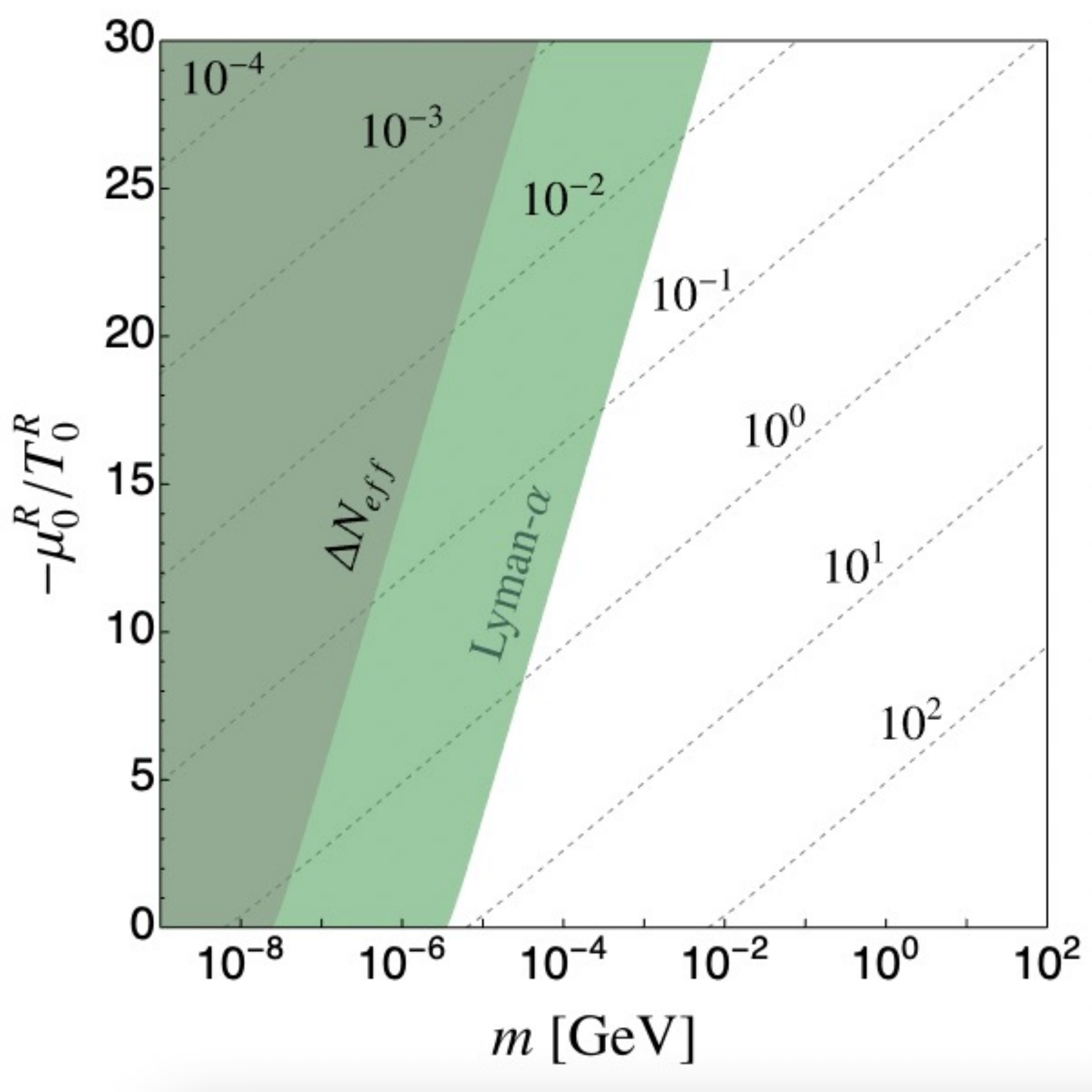}
\caption{Bounds from the effective number of relativistic species $N_{eff}$ at big-bang nucleosynthesis, $\Delta N_{eff}\leq 0.354$ \cite{Arcadi:2019oxh,PhysRevD.98.030001} (gray) and \Lya under conservative assumptions (green) on scalar singlet dark matter, as a function of the particle's primordial chemical potential and mass, where the primordial chemical potential is normalized to the comoving temperature.  
The dotted lines show contours of the value of $\TSM/T_0^R$ required for the dark matter to have the correct relic abundance (see Eq.~\eqref{eq:matching2}) with $\TSM$ being the current CMB temperature.
This plot assumes that the singlet has a vanishing quartic, so that it does not enter into chemical equilibrium. 
For a non-vanishing quartic, bounds from \Lya in the parameter space shown in this figure are only slightly weaker (as seen in Fig.~\ref{fig:scalar}), 
but a careful analysis of chemical decoupling must be done in order to check that the primordial singlet abundance has not been depleted. Note that the $N_{eff}$ bounds shown here are from only big-bang nucleosynthesis, as our dark matter particle is nonrelativistic at recombination and therefore are not constrained by CMB measurements of $N_{eff}$.}
\label{fig:Arcadi}
\end{center}
\end{figure}

When the scalar's quartic is set to zero we do not have to worry about the dark sector entering chemical equilibrium. 
For this case, in Fig.~\ref{fig:Arcadi} we present bounds on the scalar singlet from free-streaming as a function of the particle's chemical potential and mass. 
In the figure we also show bounds from the effective number of relativistic species $N_{eff}$ from big-bang nucleosynthesis as in~\cite{Arcadi:2019oxh,PhysRevD.98.030001}, 
and in dashed lines we show contours of the ratio $\TSM/T_0^R$ required to obtain the correct relic abundance, \textit{c.f.} Eq.~\eqref{eq:matching3}.
We immediately see that bounds form \Lya provide orders-of magnitude improvements over $N_{eff}$ bounds. 
When the quartic is non-zero the bounds in Fig.~\ref{fig:scalar} change little, since as discussed previously self interactions only moderately relax \Lya bounds. 
In this case, however, one must carefully determine the regions of parameter space where the field remains out of chemical equilibrium to ensure that our bounds remain valid and study the possibility of semi-relativistic decoupling.  
This is beyond the scope of this work. 

We conclude by briefly commenting on the possibility of a fermionic SIDM model with purely elastic interactions.  
In this case, an elastic and velocity-independent cross section can be obtained in the non-relativistic regime via a four-fermi interaction from a heavy mediator, as long as the mass of the mediator remains above the temperature of the dark sector. 
The simplest UV complete models leading to such interactions, are fermionic Majorana SIDM with a Yukawa coupling to a scalar heavy mediator, 
or fermionic Dirac SIDM with a heavy vector mediator.
Note that for such models the cosmology of the mediator must also be taken into account (see \textit{e.g.} \cite{Huo:2017vef}), as it easily reaches chemical equilibrium with the fermion dark matter at high temperatures.

\section{Conclusions}
\label{sec:conclusions}
In this work we studied the cosmological evolution of the dark-matter perturbations in the presence of elastic self interactions.
We included in our analysis the effect of pressure support in the growth of perturbations, 
kinetic decoupling and the period of free-streaming after decoupling.
As a result, we obtained the matter power-spectrum for SIDM for the whole range of cross sections allowed by current bounds from the Bullet Cluster, 
$0\leq \sigma/m\lesssim 1\,\cm^2/\g$.
By analyzing the amount of power suppression at small scales, 
we derived bounds from the \Lya forest using the area criterion presented in~\cite{Murgia:2017lwo}.
We found that bounds from \Lya have a slight but clear dependence on the self-interaction cross section.  
For dark matter with a vanishing primordial chemical potential, 
we found that if dark matter is kinetically decoupled by the time \Lya modes enter the horizon around $z\sim 10^6$, which happens for $\sigma/m\lesssim 10^{-5} \cm^2/\g$, dark matter is excluded if it is lighter than $m \sim 5.3 \, \keV$ ($3.5$\, \keV\, for conservative assumptions on \Lya bounds).
On the other hand, if self interactions are large, $\sigma/m \sim 1\,\cm^2/\g$, the bounds relax to $m \gtrsim 4.4 \, \keV$ ($2.95$\, \keV\, for conservative bounds).
We applied our results to one concrete dark-matter model with elastic self interactions, 
namely scalar-singlet dark matter,
and found that bounds from \Lya in this model are the most stringent for a wide range of masses or dark-matter velocity dispersions. 

We conclude by commenting on possible future directions.
One important task that remains to be done is to improve the precision of the \Lya bounds on the SIDM mass (or velocity dispersion) obtained here, by performing hydrodynamical simulations of the baryonic gas on small scales when dark matter has elastic self interactions, and comparing to flux-power spectrum data. 
\Lya bounds will improve in the future, 
as uncertainties from the thermal evolution of the interstellar medium are reduced
and more high-resolution quasar spectra are added to the data analysis~\cite{pepe2019}.
Also, to our knowledge bounds on the SIDM velocity dispersion from Milky-Way satellite counts~\cite{PhysRevD.83.043506,jethwa2018upper,kennedy2014constraining}, strong lensing~\cite{Birrer_2017,gilman2019,gilman2019_newest}, stellar streams~\cite{banik2019novel,Banik_2018_naturedm,Dalal:2020mjw}, and 
high-resolution CMB lensing measurements~\cite{Nguyen:2017zqu,Sehgal:2019ewc} have not yet been explored.

Probes of the small-scale distribution of dark matter provide 
a unique window into the dynamics of the dark sector, and test the particle nature of dark matter even if it only has gravitational interactions with the Standard Model. 
In addition, a variety of small-scale issues, such as the existence of cores in the central regions of dwarf galaxies inferred from observations of rotation curves, 
point towards the existence of interactions in the dark sector. 
The exploration of such interactions is ongoing, and may lead to fantastic discoveries in the dark sector.

\section{Acknowledgments}
We thank Riccardo Murgia and Manoj Kaplinghat for comments on the draft, 
and Riccardo Murgia for clarifications  regarding the area criterion.
We would also like to thank Neelima Sehgal, Neal Dalal and Manoj Kaplinghat for useful discussions. 
DEU is supported by Perimeter
Institute for Theoretical Physics. Research at Perimeter Institute is supported in part by the Government of
Canada through the Department of Innovation, Science
and Economic Development Canada and by the Province
of Ontario through the Ministry of Economic Development, Job Creation and Trade.
RE acknowledges support from DoE Grant DE-SC0009854, Simons Investigator in Physics Award~623940, the Heising-Simons Foundation Grant No.~79921, and the US-Israel Binational Science Foundation Grant No.~	2016153.  
DG acknowledges support from DoE Grant DE-SC0009854 and Simons Award~623940. 
ML is supported by DOE Grant DE-SC0017848.
We acknowledge the use of Perimeter Institute's Symmetry cluster.

\appendix
\section{Treatment of baryons, photons and neutrinos}
\label{app:SM}
In this appendix we discuss the implementation of baryon, photon, and neutrino perturbations in our code.
We follow the conventions of~\cite{Ma:1995ey} and work in conformal Newtonian gauge. 
Initial super-horizon conditions are taken from the same reference. 

Prior to recombination, baryons and photons are treated within the tight-coupling approximation~\cite{Gorbunov:2011zz}, 
taking $\theta_\gamma=\theta_b$.
In this period, we evolve the photon and baryon perturbations as 
\begin{eqnarray}
\nonumber 
\dot{\delta}_\gamma &=&4 \dot{\phi} -\frac{4}{3}  \theta_{b\gamma} \quad ,\\
\nonumber 
\dot{\delta}_b &=&3 \dot{\phi} -  \theta_{b\gamma}  \quad ,\\
\nonumber 
\dot{\theta}_{b\gamma} &=&k^2\psi - \bigg[  \frac{H_{\eta} R_B}{1+R_B} \theta_{b\gamma} -
\frac{3}{4} k^2 c_{b\gamma}^2 \delta_{\gamma}\bigg]
- \frac{k^2 \sigma_{\gamma}}{1+R_B} \quad ,
\end{eqnarray}
where the dots are derivatives with respect to conformal time $\eta$, $H_\eta\equiv \dot{a}/a$ and
\begin{eqnarray}
\nonumber 
R_B&=&\frac{3 \rho_b}{4 \rho_\gamma} \quad , \\
c_{b\gamma}^2 &=& \frac{1}{3(1+R_B)} \quad .
\end{eqnarray}
Photon diffusion due to the finiteness of the Thomson cross section is included  approximating the anisotropic stress by~\cite{Dodelson:2003ft}
\begin{equation}
\sigma_\gamma=
\frac{8 \theta_{b \gamma}}{27 n_e \sigma_T a} \quad ,
\end{equation}
where $n_e$ is the redshift-dependent electron number density, 
which we take from RECFAST~\cite{2000ApJS..128..407S}, and $\sigma_T$ is the Thomson cross section.
We treat photon-baryon decoupling as sudden at a recombination redshift of $z_{\textrm{rec}}=1100$. 
After decoupling, we evolve baryons and photons using 
\begin{eqnarray}
\nonumber 
\dot{\delta}_\gamma &=&4 \dot{\phi} -\frac{4}{3}  \theta_{\gamma} \quad ,\\
\nonumber 
\dot{\theta}_{\gamma} &=&
k^2 \psi +  
k^2 \bigg[\frac{1}{4} \delta_{\gamma} -\sigma_\gamma \bigg]
+ a n_e \sigma_T (\theta_b-\theta_\gamma)
\\
\nonumber
\dot{\delta}_b &=&3 \dot{\phi} -  \theta_{b}  \quad ,\\
\nonumber 
\dot{\theta}_{b} &=&k^2 \psi -  H_{\eta}\theta_{b} +
k^2c_{b}^2 \delta_{b}
+  \frac{a n_e \sigma_T}{R_B} (\theta_\gamma-\theta_b) \quad ,
\end{eqnarray}
where the baryonic speed of sound is given by
\begin{equation}
c_{b}^2=\frac{T_b}{\mu}\bigg[1-\frac{1}{3}\frac{d\ln T_b}{d\ln a}\bigg] \quad .
\label{eq:csb}
\end{equation}
In Eq.~\eqref{eq:csb}, $\mu$ is the mean molecular weight, which we approximate to $\mu=1.22\, \textrm{GeV}$~\cite{Dodelson:2003ft} and $T_b$ the baryonic temperature taken from RECFAST. To include post-recombination photon free-streaming, we set an exponential cutoff for the photon perturbations, at a photon free-streaming length, $k_{fs}=2.3\,h\,\textrm{Mpc}^{-1}$.

Regarding neutrinos, we treat them as massless, and evolve the neutrino moments according to 
\begin{eqnarray}
\nonumber 
\dot{F}_{\nu0}&=&-kF_{\nu1}+4 \dot{\phi} \\
\nonumber 
\dot{F}_{\nu1}&=&\frac{4k}{3}\bigg[ \frac{1}{4}F_{\nu0}-\frac{1}{2}F_{\nu2}+\psi\bigg] \\
\dot{F}_{\nu (l\geq 2)}&=& \frac{k}{2l+1}\big[l F_{\nu (l-1)}-(l+1)F_{\nu (l+1)}\big] \quad .
\end{eqnarray}
The neutrino density, velocity, and anisotropic stress perturbations are related to the moments by $\delta_\nu=F_{\nu0} \,,\, \theta_\nu=\frac{3}{4} k F_{\nu1}$, and $\sigma_{\nu}=\frac{1}{2} F_{\nu2}$. 
We cut the neutrino moment hierarchy at $l_{\textrm{max}}=30$ with the truncation procedure of~\cite{Ma:1995ey}.

To compute the total weighted anisotropic stress Eq.~\eqref{eq:viscositytotal} needed to obtain the conformal potential $\psi$ via Eq.~\eqref{eq:potentials}, 
in addition to the dark-matter contribution, we include both the neutrino and photon viscosities $\sigma_\gamma$ and $\sigma_\nu$. 
We validate the results of our code against CLASS in appendix \ref{eq:validation}.

\section{Validation of Boltzmann Solver}
\label{eq:validation}

In this section, we present reference validations of our Boltzmann solver. 
We compare the matter-power spectrum from our solver to that from the Cosmic Linear Anisotropy Solving System (CLASS)~\cite{Blas_2011} both for CDM and 3.5\,keV WDM.  
We use throughout the best fit cosmological parameters from~\cite{Aghanim:2018eyx}.
In all cases we take neutrinos to be massless.

\begin{figure}[ht!]
\begin{center}
\includegraphics[width=8cm]{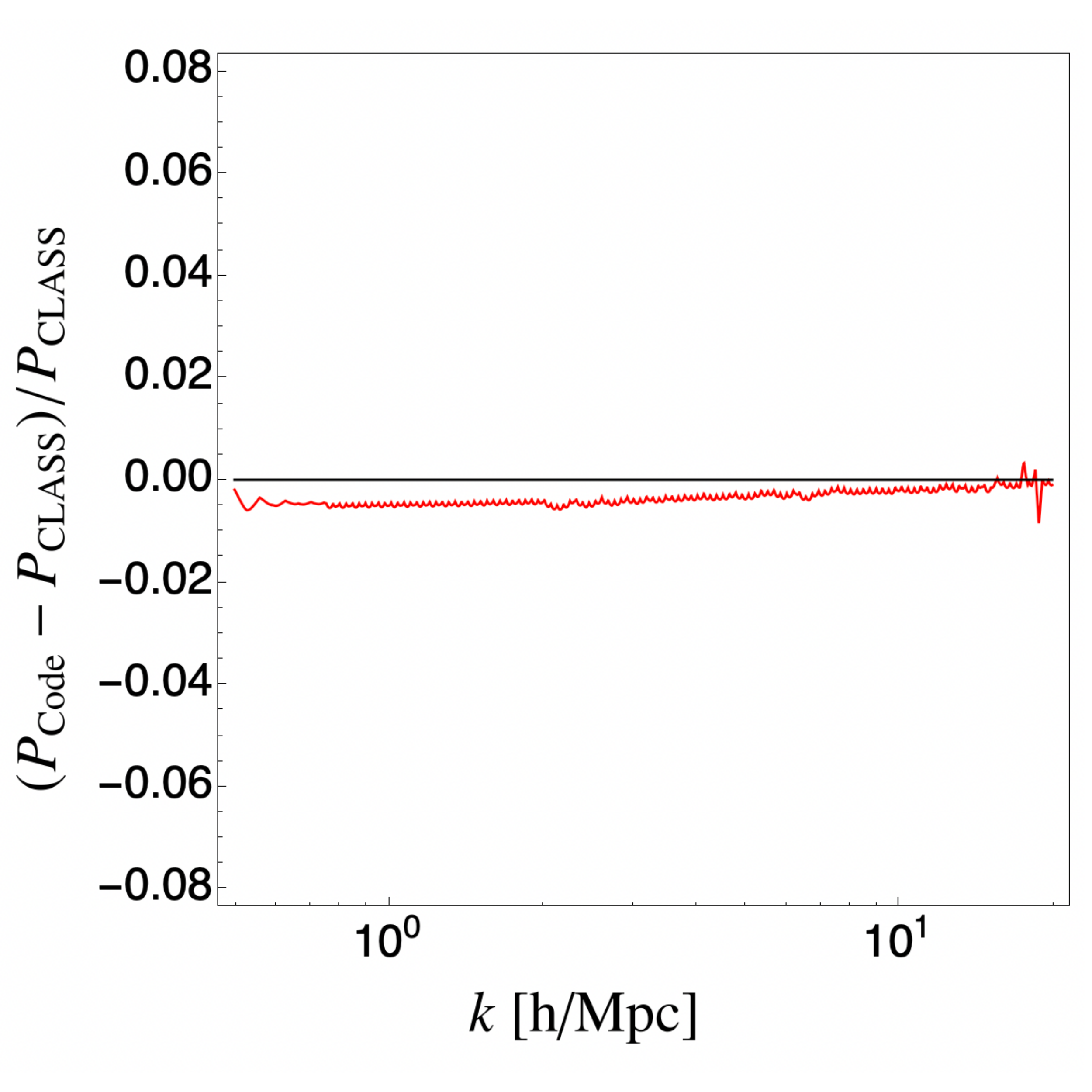}
\caption{The relative matter-power spectrum difference between the output of our Boltzmann code and a reference model from CLASS. We see that our code produces excellent agreement (within $<1\%$) for $0.5<k<20$~h/Mpc, which is our range of interest in this paper.}
\label{fig:CDMcomp}
\end{center}
\end{figure}

We compute the linear matter-power spectrum produced by our Boltzmann solver when using a massive particle with negligible velocity dispersion and vanishing self-interaction strength to simulate CDM. 
We compare this to a CDM run of CLASS. The results are given in Fig.~\ref{fig:CDMcomp}, where we plot the relative difference between the two spectra. 
We conclude that our code agrees with CLASS within the range of interest in $k$ to a precision better than $1\%$.

\begin{figure}[ht!]
\begin{center}
\includegraphics[width=8cm]{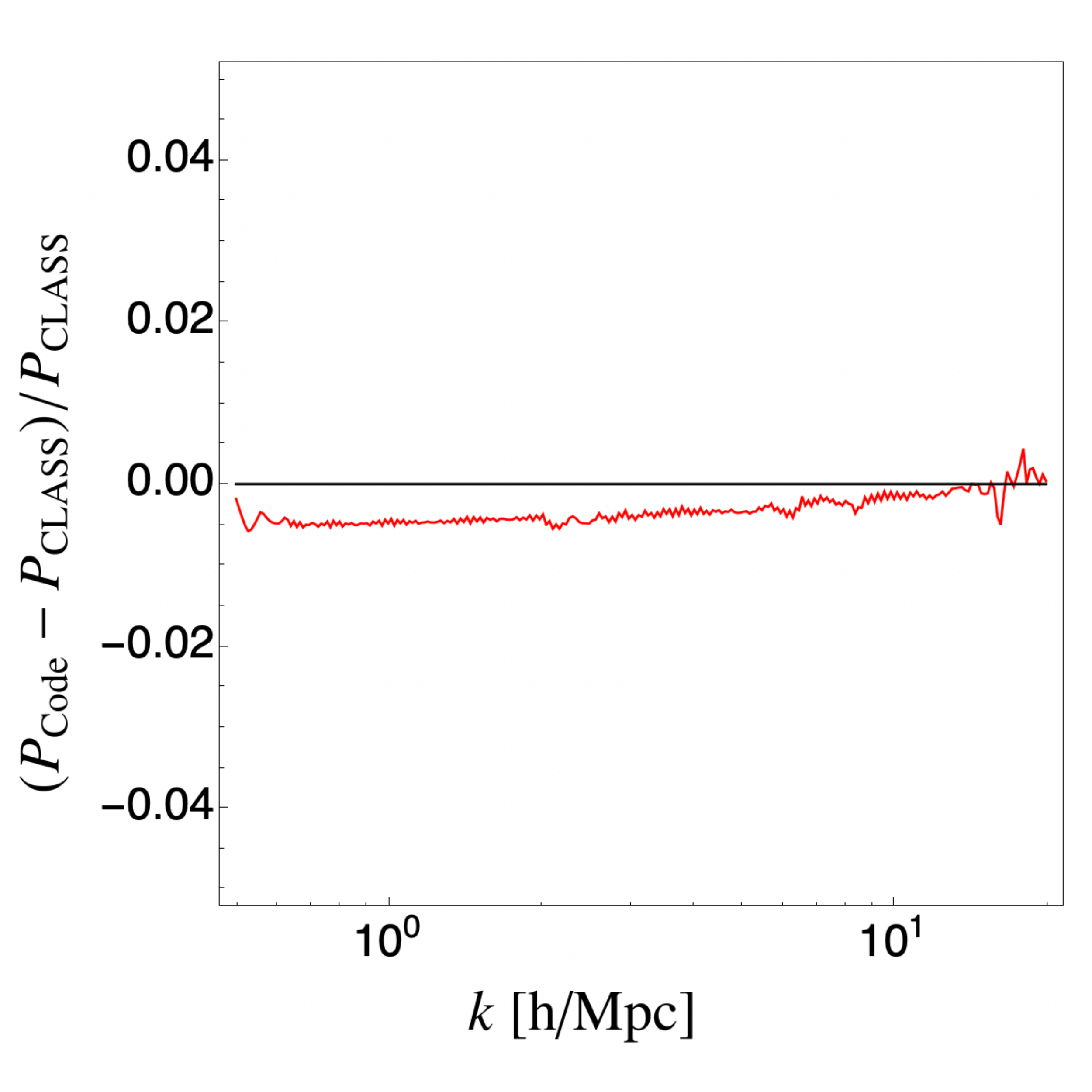}
\caption{We present the relative difference between the matter-power spectrum of our code and that of CLASS for a 3.5 keV fermionic dark-matter particle. We see that we agree with CLASS to high precision throughout the entire range of interest.}
\label{fig:WDMcomp}
\end{center}
\end{figure}

We also compare the output of our Boltzmann solver to that of CLASS for 3.5~keV WDM. 
When we run CLASS simulations for a 3.5~keV WDM candidate we do not use a fluid approximation; namely, we ask CLASS to solve the full Boltzmann hierarchy. 
The relative difference between the CLASS linear matter-power spectrum and that produced by our Boltzmann solver is shown in Fig.~\ref{fig:WDMcomp}. 
As with the CDM validation, we see our code agrees with CLASS in the $k$-range of interest with high precision. 

We take the small discrepancies in the warm-dark matter power-spectrum calculated with our code and CLASS, to be a measure of the error of our code. We have checked that these uncertainties translate into a sub-percent uncertainty in the calculation of the area estimator of reference \cite{Murgia:2017lwo} (also discussed in appendix \ref{app:areacriterion}), which is used here to set bounds from the \Lya forest.
Given that uncertainties from \Lya are expected to greatly exceed these sub-percent uncertainties (\textit{e.g.}, the area criterion has an uncertainty of about 5\%, based on an analysis of Fig.~9 from~\cite{Murgia:2018now} and a comparison of the $\delta A$ values from Table 3 from~\cite{Murgia:2017lwo}), we conclude that our code is adequate for setting \Lya bounds on the models studied throughout this work.


\section{Relaxation time approximation}
\label{app:viscosity}
The Boltzmann equation for the perturbation $\Psi$ in conformal Newtonian gauge is given by~\cite{Ma:1995ey}
\begin{equation}
\dot{\Psi} + i \frac{q}{\epsilon} (\vec{k} \hat{n}) \Psi + \frac{d \ln f_0}{d\ln q}\bigg[
\dot{\phi}-i \frac{\epsilon}{q} (\vec{k} \hat{n})\psi 
\bigg]
=
\frac{1}{f_0} C[f_0,\Psi] \quad .
\end{equation}
In general, collision terms must be calculated numerically. 
A crude estimate of such terms can be obtained in the relaxation time approximation,
where collision terms are simply approximated by their comoving collision time $\tau_{\textrm{comov}}$, 
related to the physical collision time $\tau$ of Eq.~\eqref{eq:taudef} by $\tau_{\textrm{comov}}(a)=\tau(a)/a$. 
In this approximation we simply take
\begin{equation}
\frac{1}{f_0} C[f,\Psi] \sim \frac{1}{\tau_{\textrm{comov}}} \frac{f_0 -f}{f_0}= -\frac{\Psi}{\tau_{\textrm{comov}}(a)} \quad .
\label{eq:relaxationtime}
\end{equation}
While collision terms do not affect the evolution of the first and second moment of the Boltzmann hierarchy due to conservation of particle number and kinetic energy in elastic collisions, 
they do affect the evolution of the second and higher moments. 
In particular, using the approximation Eq.~\eqref{eq:relaxationtime}, 
the evolution equation for the second moment in the presence of collision terms can be approximated to~\cite{Hannestad:2000gt}
\begin{equation}
\dot{\Psi}_2=\frac{kq}{5 \epsilon} (2\Psi_1 -3\Psi_3)- \frac{\Psi_2}{\tau_{\textrm{comov}}(a)} \quad .
\label{eq:relaxationtime2}
\end{equation} 
In the limit of large self interactions, $\tau \ll H(a)^{-1}$, so the fastest timescale in the problem
is due to collisions. 
In this case, 
we can neglect the term on the left,
leading to 
\begin{equation}
{\Psi}_2=\frac{\tau_{\textrm{comov}}(a)}{5}\frac{kq}{\epsilon} (2\Psi_1 -3\Psi_3) \quad .
\label{eq:relaxationtime3}
\end{equation} 
In addition, from the evolution equations \eqref{eq:moments},
it is clear that on scales larger than the diffusion scale, $k \tau_{\textrm{comov}} \ll 1$, 
higher moments in the hierarchy are suppressed by factors of  $k \tau_{\textrm{comov}} \ll 1$. 
In this case, we can drop moments $\Psi_{l \geq 3}$, and we get
\footnote{On scales below the diffusion scale dark-matter perturbations are suppressed by the presence of a non-zero second moment already, so it is less important to estimate higher moments of the Boltzmann hierarchy.
For a similar approximation in the context of photon diffusion damping, see~\cite{Dodelson:2003ft}.}
\begin{equation}
{\Psi}_2=\frac{2 \tau_{\textrm{comov}}(a)}{5}\frac{kq}{\epsilon} \Psi_1 \quad .
\label{eq:relaxationtime4}
\end{equation} 
With this approximation, we can now obtain the dark-matter anisotropic stress $\sigmaDM$ as a function of the velocity perturbation $\thetaDM$ using Eqns.~\eqref{eq:densityandvelocity}, \eqref{eq:viscosity} and approximating $\Psi_1$ by its adiabatic form Eq.~\eqref{eq:matchingad}.  For a relativistic species in kinetic equilibrium with $\epsilon=q$ we get
\begin{equation}
\sigmaDM= \frac{4}{15} \theta \tau_{\textrm{comov}}(a)
\label{eq:viscrel}
\end{equation}
in accordance with~\cite{Hannestad:2000gt}. 
On the other hand, deep in the non-relativistic regime, we obtain
\begin{equation}
\sigmaDM= \frac{8}{9a^2} \bigg[\frac{\rhoDM-m n_{\textrm{DM}}}{\rhoDM}\bigg]\bigg|_{a=1} \theta \tau_{\textrm{comov}}(a) \quad ,
\label{eq:viscnr}
\end{equation}
where the numerator of the term in square brackets is just the dark matter kinetic energy. 
Up to order-one numerical factors, 
the anisotropic stress in the relativistic and non-relativistic regimes match at the transition scale factor $a=\anr$,
since the ratio between the kinetic and total energy is proportional to the present velocity dispersion squared, $(\rhoDM-m n_{\textrm{DM}})/\rhoDM \sim T_0^2/m^2 \sim \anr^2$ (c.f. Eq.~\eqref{eq:anr}).
Since the comoving collision time in the non-relativistic regime grows as $a^3$, 
the right hand side of Eq.~\eqref{eq:viscnr} redshifts as $\sim\!\! a$.
Then,
in order to impose continuity of the anisotropic stress at $a=\anr$ we may simply approximate 
\begin{equation}
\sigmaDM=
\left\{
\begin{array}{cc}
\frac{4}{15} \thetaDM  \tau_{\textrm{comov}}(a) & \quad a< \anr \\
\frac{4 a}{15 \anr}  \thetaDM  \tau_{\textrm{comov}}(\anr)& \quad a > \anr \quad ,
\end{array}
\right.
\end{equation}
which is equivalent to Eq.~\eqref{eq:viscosityapprox} upon replacing $\tau_{\textrm{comov}}=\tau/a$.

\section{Matching density and velocity perturbations to moments of the phase space distribution}
\label{app:matching}

For evolving modes that are within the horizon, at kinetic decoupling we must match the density and velocity perturbations to moments of the Boltzmann equation in order to continue the simulation into the kinetically decoupled regime. 
Here we derive the matching conditions if decoupling happens in the non-relativistic regime, 
and also quote the matching conditions in the relativistic regime, which can be calculated in a similar way and have been also derived in~\cite{Ma:1995ey}.

In the non-relativistic and kinetically coupled regime, 
the phase-space distribution function Eq.~\eqref{eq:zerothorder} is locally Boltzmann-like, 
and given by
\begin{equation}
f(\vec{x}_p,t)=\frac{n(\vec{x}_p)}{\big[ 2\pi m T(\vec{x}_p)\big]^{3/2} } \exp\bigg[-\frac{(\vec{p}- m \vec{v}(\vec{x}_p,t))^2}{2  m T(\vec{x}_p)} \bigg]\,,
\end{equation}
where $\vec{x}_p,\vec{p}$ are physical spatial and momenta, 
$\vec{v}$ is a gas bulk velocity perturbation field,
$m$ the dark-matter mass,
and $n(\vec{x}_p,t)$, $T(\vec{x}_p,t)$ are the local number density and temperature of the dark-matter gas.
They are given by constant homogeneous background values plus perturbations,
\begin{eqnarray}
\nonumber
n(\vec{x}_p,t)&=&\bar{n}(t)+\delta n (\vec{x}_p,t) \\
T(\vec{x}_p,t)&=&\bar{T}(t)+\delta T (\vec{x}_p,t) \,.
\end{eqnarray}
Expanding the distribution function $f$ to lowest order in terms of the density, temperature, and bulk velocity perturbation fields,
we obtain the inhomogeneous piece of the distribution, $\Psi(\vec{x}_p,t)$ (c.f.~Eqn.~\eqref{eq:zerothorder}),
\begin{equation}
\Psi(\vec{x}_p,t)= \frac{\delta n(\vec{x}_p,t)}{\bar{n}(t)} 
+\bigg[ \frac{p^2}{2  m \bar{T}(t)^2}-\frac{3}{2\bar{T}(t)}\bigg] \delta T(\vec{x}_p,t)
+ \frac{\vec{v}(\vec{x}_p,t)\cdot \vec{p} }{\bar{T}(t)}\,.
\label{eq:psi}
\end{equation}
Assuming adiabaticity, 
the number density and temperature perturbations for an ideal monoatomic gas are related by
\begin{equation}
\delta T = \frac{2}{3} \frac{\delta n}{\bar{n}} \bar{T} \quad .
\label{eq:adiabaticity}
\end{equation}
In addition, 
only longitudinal velocity perturbations propagate in the gas~\cite{Gorbunov:2011zzc}, 
so moving into comoving Fourier space we may take 
\begin{equation}
\vec{v}(\vec{k})=\hat{k} v_l(\vec{k})\,,
\label{eq:bulkvel}
\end{equation}
with $\hat{k}$ being the comoving spatial Fourier mode vector of unit norm and $v_l(\vec{k})$ a longitudinal scalar velocity perturbation field. 
Using Eqns.~\eqref{eq:adiabaticity} and \eqref{eq:bulkvel} in \eqref{eq:psi}, 
and $d \ln f_0/d\ln p=-p^2/m\bar{T}$, we obtain
\begin{equation}
\Psi(\vec{k},\vec{p},t)= -  \frac{1}{3}\frac{\delta n (\vec{k},t)}{\bar{n}(t)} \frac{d\ln f_0}{d \ln p}  - \frac{(\hat{k}\cdot{\hat{p}}) m  v_l(\vec{k},t)}{p}  \frac{d\ln f_0}{d \ln p} \,.
\label{eq:psi2}
\end{equation}
To provide contact with the notation in~\cite{Ma:1995ey} used in the main body of the text, 
we note that the physical velocity and number density perturbations are related with $\deltaDM$ and $\thetaDM$ by\footnote{The relation between $\thetaDM$ and the physical velocity perturbation $v_l$ is found by matching the energy momentum tensor in~\cite{Ma:1995ey}  to the one of an ideal fluid with bulk velocity.}
\begin{eqnarray}
\nonumber
\frac{\delta n (\vec{k},t)}{\bar{n}(t)} &=& \deltaDM(\vec{k},t) \quad ,\\
v_l(\vec{k},t) &=& -\frac{i\theta(\vec{k},t)}{k} \quad .
\label{eq:notationmatch}
\end{eqnarray}
Using \eqref{eq:notationmatch} and expressing physical momenta in terms of comoving momenta, $p=q/a$, we get
\begin{equation}
\Psi(\vec{k},\vec{q},t)= -  \frac{\deltaDM(\vec{k},t)}{3}\frac{d\ln f_0}{d \ln q}   + \frac{i  (\hat{k}\cdot{\hat{q}}) \theta(\vec{k},t) a m}{qk}  \frac{d\ln f_0}{d \ln q} \,.
\label{eq:psi3}
\end{equation}
Now, the Boltzmann Legendre mode decomposition is given by
\begin{equation}
\Psi_l(\vec{k},q,t)=\int_{-1}^{1} d\mu \Psi(\vec{k},\mu,q,t)  \frac{P_l(\mu) }{2(-i)^l}\,,
\label{eq:modes}
\end{equation}
where $\mu=\hat{k}\cdot \hat{q}$ and $P_l$ are Legendre polynomials.
Using Eqns.~\eqref{eq:psi3} and \eqref{eq:modes} we get
\begin{eqnarray}
\nonumber \Psi_0&=& -\frac{\deltaDM}{3}   \frac{d\ln f_0}{d \ln q} 
\\
\Psi_1&=& -\frac{\theta a m}{3qk}  \frac{d\ln f_0}{d \ln q}  \quad ,
\label{eq:matching1}
\end{eqnarray}
with higher moments vanishing, 
which is consistent with the assumptions of treating the fluid as ideal and with an adiabatic evolution.
Deviations from the ideal fluid are obtained from anisotropic stress due to dark-matter diffusion, as discussed in appendix \ref{app:viscosity}.
In order to include these effects, 
we also match the second moment according to Eq.~\eqref{eq:relaxationtime4}.
Combining Eq.~\eqref{eq:relaxationtime4} and \eqref{eq:matching1}, 
we get
\begin{equation}
\Psi_2 =  -\frac{2 \theta \tau_{\textrm{comov}}(a)}{15}   \frac{d\ln f_0}{d \ln q}  \quad .
\end{equation}

In the relativistic regime, on the other hand, the derivation of the matching conditions is similar.
The matching conditions are
\begin{eqnarray}
\nonumber \Psi_0&=& -\frac{\deltaDM}{4}   \frac{d\ln f_0}{d \ln q} \\
\nonumber \Psi_1&=& -\frac{\theta }{3k}  \frac{d\ln f_0}{d \ln q}  \\
\Psi_2 &=&  -\frac{2 \theta  \tau_{\textrm{comov}}(a)}{15}   \frac{d\ln f_0}{d \ln q}  \quad .
\label{eq:matchingrel}
\end{eqnarray}
The matching conditions Eqns.~\eqref{eq:matching1}-\eqref{eq:matchingrel} are equivalent to Eq.~\eqref{eq:matchingad} upon replacing $\tau_{\textrm{comov}}=H_{\eta}^{-1}$ at decoupling.
We also note that the matching depends only on adiabaticity. 
Thus, the same matching conditions can be used for frozen modes that are outside the horizon, prior to horizon entry, as in \cite{Ma:1995ey}.

\section{The Area Criterion}
\label{app:areacriterion}

In the area criterion, an ``area'' estimator of the power suppression $A$ is defined as~\cite{Murgia:2017lwo}
\begin{equation}
A \equiv \int_{k_{\textrm{min}}}^{k_{\textrm{max}}} dk \, \frac{P_{\textrm{1D}}(k)}{P_{\textrm{1D}}^\textrm{CDM}(k)}\,, 
\label{eq:areaestimator}
\end{equation}
where $k_{\textrm{min}}=0.5 \, h/\Mpc$ and $k_{\textrm{max}}=20 \, h/\Mpc$ set the  range of comoving momenta probed by Lyman-$\alpha$,
and $P_\textrm{1D}(k)$ is the one-dimensional matter-power spectrum defined as
\begin{equation}
P_\textrm{1D}(k) \equiv \frac{1}{2\pi} \int_k^{\infty} dk' k' P(k')  \quad .
\label{eq:1DPS}
\end{equation}
In Eq.~\eqref{eq:1DPS}, $P(k)$ is the present matter-power spectrum calculated using linear theory.

In order to set bounds on dark-matter models, in the area criterion one first calculates the relative difference between the area estimator $A_{\textrm{CDM}}$ of the $\Lambda \textrm{CDM}$ model, 
and the estimator of a reference model $A_r$, which is known from detailed simulations to be at the $95\%$ exclusion boundary,
\begin{equation}
\delta A_r = \frac{A_{\textrm{CDM}}-A_r}{A_{\textrm{CDM}}}\,.
\label{eq:relativeestimator}
\end{equation}
Following~\cite{Irsic:2017ixq}, here we take the reference model to be  warm dark matter with a mass of $5.3 \,\keV$.
Using our Boltzmann code, we obtain the power-spectrum of warm dark matter and get
\begin{equation}
\delta A_r = 0.049  \quad .
\label{eq:relativereference}
\end{equation}
Finally, in order to set bounds on a given dark-matter model, 
the area estimator Eq.~\eqref{eq:areaestimator} for that model is obtained, 
and from there the relative estimator \eqref{eq:relativeestimator} is calculated. 
If such relative estimator is larger than the reference relative estimator in Eq.~\eqref{eq:relativereference}, the model is excluded at the $95\%$ confidence level. 
For conservative bounds, taking the excluded WDM mass to be $3.5 \,\keV$, $\delta A_r =0.135$.

The accuracy of the area criterion has been validated against a variety of dark-matter models in~\cite{PhysRevD.98.083540}, where the authors found excellent agreement between exclusion of models obtained by performing complete numerical simulations of flux power spectra and statistical analyses of Lyman-$\alpha$ data and exclusion of models using the much simpler area criterion. 
Further tests on the robustness of the criterion have been performed in \cite{DEramo:2020gpr}.


\bibliography{siwdm}

\end{document}